\documentclass[english]{revtex4-1}
\usepackage[LGR,LGR,T1]{fontenc}
\usepackage[latin9]{inputenc}
\usepackage{color}
\usepackage{float}
\usepackage{textcomp}
\usepackage{mathrsfs}
\usepackage{bm}
\usepackage{amsmath}
\usepackage{graphicx}

\makeatletter

\DeclareRobustCommand{\greektext}{%
  \fontencoding{LGR}\selectfont\def\encodingdefault{LGR}}
\DeclareRobustCommand{\textgreek}[1]{\leavevmode{\greektext #1}}
\ProvideTextCommand{\~}{LGR}[1]{\char126#1}

\usepackage{fleqn}
\usepackage{epsfig}\usepackage{slashed}
\usepackage{bbold}

\def\onehalf{{1 \over 2}}

\def\ps {\not\!p}
\def\ks {\not\!k}
\def\qs {\not\!q}

\def\br{\begin{eqnarray}}
\def\er{\end{eqnarray}}

\usepackage{babel}

\usepackage{babel}

\usepackage{babel}

\makeatother

\usepackage{babel}
\begin{document}

\title{\textsc{\Large{}{}Weak pion-production and the second resonance
region}}

\author{{\normalsize{}{}D.F. Tamayo Agudelo$^{1}$,A. Mariano$^{2,3}$,
and $^{1}$D.E. Jaramillo Arango}}

\affiliation{\textrm{$^{1}$Facultad de C. Exactas y Naturales Universidad de
Antioquia, Ciudad Universitaria: Calle 67 N$^{0}$ 53-108 Bloque 6
Oficina 105, Medellin, Colombia Instituto de F�sica,~~ }~\\
 \textrm{$^{2}$Facultad de Ciencias Exactas Universidad Nacional
de La Plata,C.C. 67,1900 La Plata,Argentina. ~~}~\\
 \textrm{ $^{3}$Instituto de F�sica La Plata CONICET, diagonal 113
y 63, 1900 La Plata, Argentina. ~~}~~\\
 ~~\\
 }
\begin{abstract}
{\normalsize{}{}In this work we report the calculation of the total
cross section for pion-production by the dispersion of neutrinos on
nucleons. We use a consistent formalism for the intermediate resonance
states of spin $\frac{3}{2}$, taking care on the conditions imposed
by the invariance on contact transformations and using the Sachs parametrization
for the form factors. In addition, we incorporate states in the second
resonance region up to 1.6 GeV and implement different approaches
for all the dressed resonance propagators. In addition to the $\Delta(1232)$,
we include the $N^{*}(1440)$,$N^{*}(1535)$and $N^{*}(1520)$ resonance
contributions, and it is shown that this inclusion improve the description
of the total cross section regards a model where only the $\Delta$(1230)
resonance is considered. Also results for antineutrinos are properly
described.}{\normalsize \par}

\smallskip{}
\end{abstract}
\maketitle

\section{INTRODUCTION}

In the standard model, neutrinos are massless, but experimental evidence
shows that although small, it is not zero. This increases interest
in their study as it leads to flavor oscillations, mixture of angles
between mass states, violation of quantum numbers, among others. The
detection of neutrino masses is the first evidence of physics beyond
the standard model. Neutrino physics has been one of the most studied
topics in recent years for particle physics. Now, as it is known that
neutrinos are massive particles that can oscillate (changing flavor),
it is essential to know precisely the cross section in the interaction
of the neutrino with nucleons or with a nucleus in the detector. The
interaction of neutrinos with nuclei and nucleons have received considerable
attention in recent years, stimulated by the needs in the analysis
of neutrino\textsf{\large{}{}}\textsf{ }experiments giving information
about the probability of oscillation\textsf{{}.} There are several
processes for the study of the interaction of neutrinos with nucleons.
The dispersion of neutrinos by nucleons can be quasielastic or inelastic
producing additional pions together the nucleon in charged current
(CC) and neutral current (NC) interactions \cite{Llewe72}. CC quasi-elastic
(QE) interaction of neutrinos and antineutrinos with nucleons involves
the processes $\nu_{\ell}+n\rightarrow p+\ell^{-}$and $\bar{\nu}_{\ell}+p\rightarrow n+\ell^{+}$
respectively where\textsf{{} }$\ell=e$, $\nu$, $\tau$\textsf{{},
}and it is used to detect the arriving of neutrinos or antineutrinos
to the detectors. The following process in importance to be considered
is the CC single pion production (SPP) $\nu_{\ell}+N\rightarrow\ell^{-}+N'\pi$
and $\bar{\nu}_{\ell}+N\rightarrow\ell^{+}+N'\pi$ , where $N,N'=$
$p$, $n$ or the NC one $\nu_{\ell}+N\rightarrow\nu'_{\ell}+N'\pi$
and $\bar{\nu}_{\ell}+N\rightarrow\bar{\nu}'_{\ell}+N'\pi$ .

A good understanding of SPP by neutrinos with few-GeV energies is
important for current and future oscillation experiments, where pion
production is either a signal process when scattering cross sections
are analyzed, or a large background for analyses which select QE events.
At these energies, the dominant production mechanism is via the production
and subsequent decay of hadronic resonances.

The axial form factor(FF) for pion production on free nucleons cannot
be constrained by electron scattering data, used normally to get the
vector FF, so it relies upon data from Argonne National Laboratory's
$12$ ft bubble chamber (ANL)\cite{Rad82} and Brookhaven National
Laboratory's $7$ ft bubble chamber (BNL)\cite{Kitagi86}. The ANL
neutrino beam was produced by focusing 12.4 GeV protons onto a beryllium
target. Two magnetic horns were used to focus the positive pions produced
by the primary beam in the direction of the bubble chamber, these
secondary particles decayed to produce a predominantly $\nu_{\mu}$
peaked at $\sim0.5$ GeV. The BNL neutrino beam was produced by focusing
29 GeV protons on a sapphire target, with a similar two horn design
to focus the secondary particles. The BNL $\nu_{\mu}$ beam had a
higher peak energy of $\sim1.2$ GeV, and was broader than the ANL
beam. These experiments will be referenced below in the results section.
These datasets differed in normalization by 30\textendash 40$\%$
for the leading pion production process $\nu_{\mu}p\rightarrow\mu^{-}p\pi^{+}$,
which conduced to large uncertainties in the predictions for oscillation
experiments.

It has long been suspected that the discrepancy between ANL and BNL
was due to an issue with the normalization of the flux prediction
from one or both experiments, and it has been shown by other authors
that their published results are consistent within the experimental
uncertainties provided\cite{Graczyk09,Graczyk14}. In Ref. \cite{wilki14},
was presented a method for removing flux normalization uncertainties
from the ANL and BNL $\nu_{\mu}p\rightarrow\mu^{-}p\pi^{+}$ measurements
by taking ratios with charged current quasielastic (CCQE) event rates
in which the normalization cancels. Then, it was obtained a measurement
of $\nu_{\mu}p\rightarrow\mu^{-}p\pi^{+}$by multiplying the ratio
by an independent measurement of CCQE (which is wellknown for nucleon
targets). Using this technique, they found good agreement between
the ANL and BNL $\nu_{\mu}p\rightarrow\mu^{-}p\pi^{+}$datasets. Later,
they extend that method to include the subdominant $\nu_{\mu}n\rightarrow\mu^{-}p\pi^{0}$
and $\nu_{\mu}n\rightarrow\mu^{-}n\pi^{+}$channels\cite{Rodriguez16}.
This is one of the reasons encourage us to return to the calculation
of SPP neutrino-nucleon cross sections. The other reason is that there
are many models to describe this process that fail in several aspects
namely:\\
 i)There are problems from the formal point of view. Since the pion
emission source are excitation and decay of resonances, and many of
them are of spin $\frac{3}{2}$, we must keep amplitudes invariant
by contact transformations (see below). These transformations change
the amount of the spin-$\frac{1}{2}$ spurious contribution in the
field that are present by construction. Many works keep the simpler
forms of the free and interaction Lagrangians, and the amplitude lacks
the mentioned invariance\\
 ii)In addition to the resonances pole contribution(normally referred
as resonant terms) to the amplitude, we have background terms coming
form cross resonance contributions and non-resonance origin (called
usually non resonant terms). Many works do not consider the interference
between resonant and background contributions and really it is very
important to describe the data.\\
 iii)Another models detach the decay process for the resonance out
of the whole weak production amplitude. However, resonances are nonperturbative
phenomena associated to the pole of the S-matrix amplitude and one
can not detach its production from its decay mechanisms, omitting
the details in the propagation.

In this work we calculate the SPP cross section, where we use a consistent
formalism for the intermediate resonance states of spin $\frac{3}{2}$.
The $\Delta(1232)$ will be described within the Complex Mass Scheme
(CMS), obtained from its dressed propagator (see below). In addition,
we incorporate states in the second resonance region which includes
the $N^{*}(1440),N^{*}(1535),N^{*}(1520)$ resonances treated within
a constant width approach . This work is organized as follows: In
Section II, summarize the general description of weak interactions
and the SPP cross section. In Section III we will introduce the formalism
of Rarita -Schwinger for spin-$\frac{3}{2}$ particles, the dressed
propagator for them and propose the consistent form of the vertex
and the propagator. Also, we introduce the formalism for the $\Delta(1232)$
and resonances of the second region ($N^{*}(1440),N^{*}(1520),N^{*}(1535)$).
In Section IV we show the results obtained with our model in the different
regimes of the final $W_{\pi N}$ invariant mass. Finally in Section
VI we summarize our conclusions.

\section{Neutrino-nucleon scattering}

The CC interaction (we omit other contributions) between a neutrino
and a hadron is obtained from the weak Lagrangian

\begin{eqnarray}
\mathscr{\mathcal{L}}_{CC} & = & \frac{-g}{2\sqrt{2}}\left(J_{lCC}^{\mu}W_{\mu}+J_{hCC}^{\mu}\sqrt{2}\left(\bm{\tau}\mbox{or }\bm{T}^{\dagger}\right)\cdot\bm{W}+h.c.\right)\label{eq:weakLagrangian}
\end{eqnarray}
being the leptonic and hadronic currents respectively

\begin{eqnarray}
J_{lCC}^{\mu} & = & \sum_{l}\bar{\psi_{l}}\gamma^{\mu}(1-\gamma_{5})\psi_{\nu_{l}},\nonumber \\
J_{hCC}^{\mu} & = & \bar{\psi}_{h'}(V^{\mu}-A^{\mu})\psi_{h},\label{eq:hadroncurrent}
\end{eqnarray}
where the isospin operator $\bm{\tau}$, the ${\bf T^{\dagger}}$
$N\rightarrow R(I=\frac{3}{2})$ excitation one, and the isospin wave
functions for the bosons $\bm{W}_{\pm}$( equal to the pions $\boldsymbol{\phi}_{\pm}$),
the nucleons $N=p,n$ and resonances R, are defined in the appendix
A. Finally we have the $W$ propagator

\begin{eqnarray*}
D_{\mu\nu}(p) & = & \frac{-g_{\mu\nu}+\frac{p_{\mu}p_{\nu}}{m_{W}^{2}}}{p^{2}-m_{W}^{2}}\approx\frac{g_{\mu\nu}}{m_{W}^{2}},
\end{eqnarray*}
being $\frac{g_{\mu\nu}}{m_{W^{2}}}$ the the high mass limit since
usually $p^{2}\ll m_{W}^{2}$.

In this work we analyze the CC $\nu N\rightarrow\mu^{-}N'\pi$ and
$\bar{\nu}N\rightarrow\mu^{+}N'\pi$ modes. The total amplitude $\mathcal{M}$
can be expressed from the Lagrangian (\ref{eq:weakLagrangian}) as
(spin and isospin indexes omitted) 
\begin{eqnarray}
\mathcal{M} & = & \frac{ig^{2}}{\left(2\sqrt{2}\right)^{2}}\bar{u}(p_{\mu})(-)i\gamma^{\lambda}(1-\gamma_{_{5}})u(p_{\nu})\frac{ig_{\lambda\lambda'}}{m_{W}^{2}}V_{ud}\bar{u}(p')\mathcal{O}^{\lambda'}(p',k,p,q)u(p),\label{eq:amplitude}
\end{eqnarray}
where$\frac{g^{2}}{\left(2\sqrt{2}\right)^{2}}=\frac{G_{F}^{2}}{\sqrt{2}}$,
$G_{F}=1.16637\times10^{-5}GeV{}^{-2}$ , $|V_{ud}|=0.9740$, and
the 4-momenta are defined as

\begin{eqnarray*}
p_{\nu}=(E_{\nu},{\bf p}_{\nu}),\hspace{0.5cm}p_{\mu}=(E_{\mu},{\bf p}_{\mu}),\hspace{0.5cm}\text{k}=(E_{\pi},{\bf k}),\hspace{0.5cm}p=(E_{N},{\bf p}),\qquad p'=(E_{_{N'}},{\bf p}')
\end{eqnarray*}
with $E_{i}=\sqrt{|{\bf p}_{i}|^{2}+m_{i}^{2}}$, we set $m_{\nu}=0$,
and $\mathcal{O}^{\lambda}$ is the vertex generated by $J_{hCC}^{\mu}$
in the Lagrangian (\ref{eq:weakLagrangian}). We will built $\bar{u}O^{\lambda}u$
for the $\nu(\bar{\text{\ensuremath{\nu}}})N\rightarrow\mu N'\pi$
process with the contributions shown in Fig.1. Clearly, all the Feynman
graphs do not necessarily contribute to each of channels as we will
analyze.

\begin{figure}[h]
\includegraphics[scale=0.3]{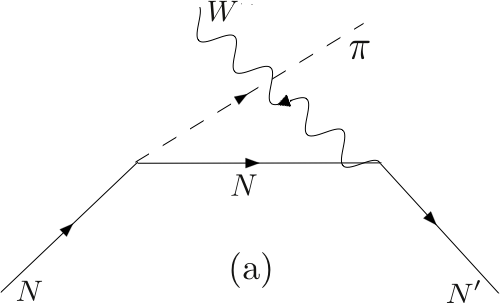}\includegraphics[scale=0.3]{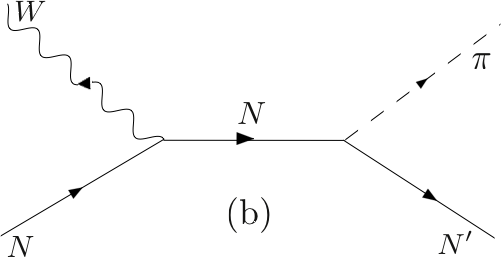}

\includegraphics[scale=0.3]{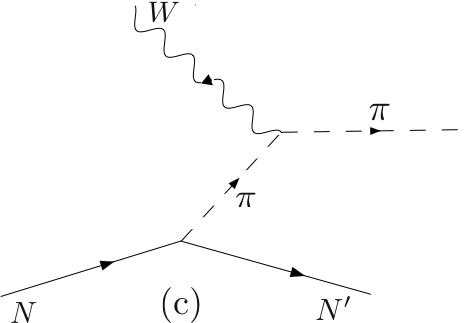}\includegraphics[scale=0.3]{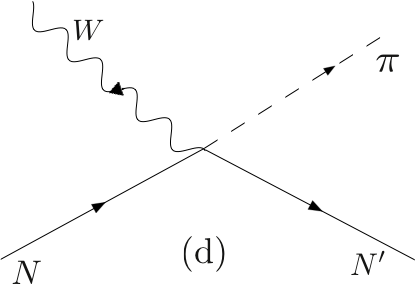}

\includegraphics[scale=0.3]{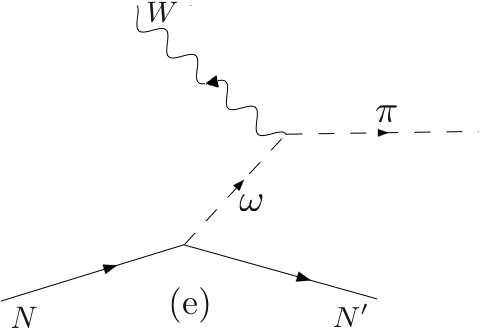}\includegraphics[scale=0.3]{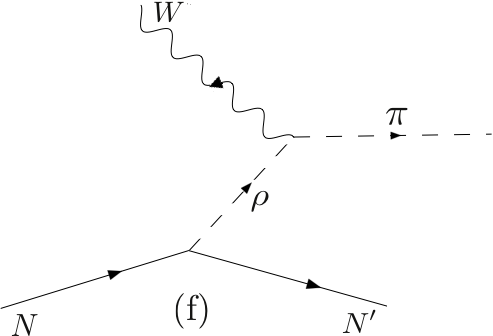}

\includegraphics[scale=0.3]{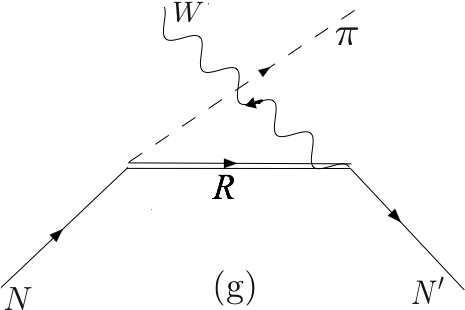}\includegraphics[scale=0.3]{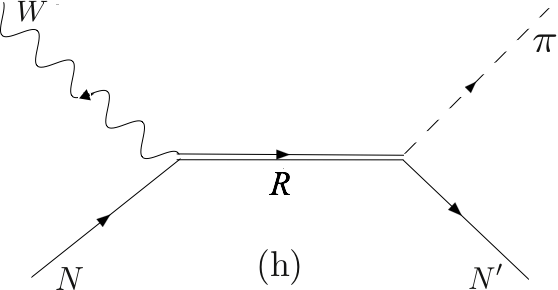}

\caption{Contributions to the scattering amplitude for the process $\nu(\bar{\text{\ensuremath{\nu}}})N\rightarrow\mu N'\pi$.
Fig (a)-(g) are the background (B) contributions. Fig (h) are the
pole resonant contributions (R). $R$ in the Figure indicates any
of the intermediate considered resonances.}
\end{figure}

We assume a tree-level hadronic amplitudes contributing to the so
called background (B) that encloses the nucleon Born terms (Fig. 1(a)-(b)),
the meson exchange amplitudes(including the contact term) (Figs. 1(c)-(f))
and the resonance($R)$-crossed term (Fig. 1(g)); the pole resonant
contribution (R) is shown in Fig. 1(h) and as we will show the resonance
acquires a width (i.e we don't have  only a tree-label amplitude)
that avoids the singularity in the propagator by a dressing it at
different levels of approximation. In this way, we split the hadronic
operator $\mathcal{O}^{\lambda}$ involved in the hadronic amplitude
in Eq.(\ref{eq:amplitude}) as

\begin{eqnarray}
\mathcal{O}^{\lambda} & = & \mathcal{O}_{\text{B}}+\mathcal{O}_{\text{R}},\label{eq:AmplitudB+R}
\end{eqnarray}
where as will be seen $\mathcal{O}_{B,R}$ are built through the Feynman
rules obtained from the different effective Lagrangians.

\textcolor{black}{The total cross}\textcolor{red}{{} }section for weak
SPP in terms of the $\nu N$ center mass (CM) variables for convenience,
will be calculated from (we take $\boldsymbol{p}_{\nu}=E_{\nu}\hat{\boldsymbol{k}}$
along the $Z$axis) 
\begin{eqnarray}
\sigma(E_{_{\nu}}^{\text{CM}}) & = & \frac{m_{_{\mu}}m_{_{N}}^{2}}{(2\pi)^{4}E_{_{\nu}}^{\text{CM}}\sqrt{s}}\int\limits _{E_{\mu}^{^{-}}}^{E_{\mu}^{^{+}}}dE_{_{\mu}}^{\text{CM}}\int\limits _{E_{\pi}^{^{-}}}^{E_{\pi}^{^{+}}}dE_{_{\pi}}^{\text{CM}}\int\limits _{-1}^{+1}d\cos\theta\int\limits _{0}^{2\pi}d\eta\frac{1}{16}\sum\limits _{\text{spin}}|\mathcal{M}|^{2},\label{eq:Totalcross}
\end{eqnarray}
where $\sqrt{s}=\sqrt{(p_{\nu}+p_{N})^{2}}=E_{_{\nu}}^{\text{CM}}+E_{_{N}}^{\text{CM}}$,
the angular variable come from the integration elements $d\Omega_{\mu}=d\cos\theta d\phi$
and $d\Omega_{\pi}=dcos\xi d\eta$ ($d\phi$ integration gives a factor
$2\pi$ and $\cos\xi$ is fixed by energy conservation) and

\begin{eqnarray}
E_{_{\mu}}^{-} & = & m_{_{\mu}},\:E_{_{\mu}}^{^{+}}=\frac{s+m_{_{\mu}}^{2}-(m_{_{N}}+m_{_{\pi}})^{2}}{2(E_{_{\nu}}^{CM}+E_{_{N}}^{\text{CM}})},\nonumber \\
E_{_{\pi}}^{\pm} & = & \frac{(\sqrt{2}-E_{_{\mu}}^{CM})(s-2\sqrt{s}E_{_{\mu}}^{CM}-\Delta_{m}^{2})\pm A\sqrt{(E_{_{\mu}}^{CM})^{2}-m_{\mu}^{2}}}{2(s-2\sqrt{s}E_{_{\mu}}^{CM}+m_{_{\mu}}^{2})},\label{eq:limits0}
\end{eqnarray}
with 
\begin{eqnarray}
A & = & \sqrt{(s-2\sqrt{s}E_{_{\mu}}^{CM}-\Delta_{m}^{2})^{2}-4m_{_{\pi}}^{2}(s-2\sqrt{s}E_{_{\mu}}^{CM}+m_{_{\mu}}^{2})},\nonumber \\
\Delta_{m}^{2} & = & m_{_{N}}^{2}-m_{_{\mu}}^{2}-m_{_{\pi}}^{2}.\label{eq:limits}
\end{eqnarray}
The neutrino energy CM energy is related with the laboratory one as

\begin{eqnarray}
E_{_{\nu}}^{CM} & =\frac{m_{N}E_{\text{\ensuremath{\nu}}}^{Lab}}{\sqrt{s}} & \frac{m_{_{N}}E_{_{\nu}}^{Lab}}{\sqrt{2E_{_{\nu}}^{Lab}m_{_{N}}+m_{_{N}}^{2}}}.\label{eq:EcmLab}
\end{eqnarray}
It is wellknown that the hadronic currents $J_{hCC}^{\lambda}$ have
a vector-axial structure $J_{hCC}^{\lambda}\equiv V^{\lambda}-A^{\lambda}$.
In terms of the vector current, the electromagnetic one is written
as $J_{\text{elec}}^{\lambda}=V_{\text{isoscalar}}^{\lambda}+V_{3}^{\lambda}$
($V_{3}^{\lambda}=\tau_{3}\frac{\gamma^{\lambda}}{2}$ for a nucleon
or R($I=\frac{1}{2})$, and $V^{\lambda}T_{3}^{\dagger}$ for the
$R(I=\frac{3}{2})$) and the weak vector CC is obtained through the
CVC hypothesis ($\tau_{3},T_{3}^{\dagger}\rightarrow\sqrt{2}\tau_{\pm},\sqrt{2}T_{\pm}^{\dagger}$)
as $V_{\pm}^{\lambda}\equiv\mp(V_{1}^{\lambda}\pm iV_{2}^{\lambda})=\sqrt{2}\bm{V}\cdot\bm{W}_{\pm}$.
In the same way it is also possible to get information on different
contributions to the vector current in $\mathcal{\mathcal{M_{B}}}$
as the effective $WM\rightarrow\pi'$ (with $M\equiv\pi,\omega$)
and the contact $NW\pi\rightarrow N'\pi'$ vector vertexes. Here the
FF are again obtained assuming CVC from the electromagnetic $\gamma M\rightarrow\pi'$
and $\gamma N\pi\rightarrow N'\pi'$ vertexes obtained through the
corresponding effective interaction Lagrangians, making the replacement
$(\bm{\Phi}_{\pi}^{'*}\times\bm{\Phi}_{\pi})_{3}\rightarrow\mp[(\bm{\Phi}_{\pi}^{'*}\times\bm{\Phi}_{\pi})_{1}\pm i(\bm{\Phi_{\pi}}^{'*}\times\bm{\Phi}_{\pi})_{2}]=\sqrt{2}(\bm{\Phi}_{\pi}^{'*}\times\bm{\Phi}_{\pi})_{\pm}=\surd2(\bm{\Phi}_{\pi}^{'*}\times\bm{\Phi}_{\pi})\cdot\bm{W}_{\pm}$
(the same is valid for the contact vertex changing $\boldsymbol{\Phi}_{\pi}\rightarrow\tau$)
for $M=\pi$ or $\boldsymbol{\Phi}_{\pi_{3}}^{'}\rightarrow\sqrt{2}\bm{\Phi}_{\pi}'\cdot\bm{W}_{\pm}$
for $M=\omega$. The $\rho$-exchange in the $\rho W\text{\textrightarrow}\pi$
vertex does not contribute to the vector current since the $\rho-\pi$
current is isoscalar. We assume a phenomenological $\rho W\rightarrow\pi$
axial vertex, where the isospin factor is the same as in the $\pi'W\rightarrow\pi$
vector case.

\section{Resonances }

In this section we will show the resonance Lagrangians to built $\mathcal{O}_{R}$
since those used to get $\mathcal{O}_{B}$ are more general and will
be referenced from the appendix B.

\subsection{Spin $\frac{3}{2}$ resonances}

\subsubsection{$\Delta$(1232) resonance}

This $IJ^{\pi}=\frac{3}{2},\frac{3}{2}^{+}$ positive parity resonance
has a three quark orbital momentum and spin $L=0,S=\frac{3}{2}$ and\textsf{\large{}{}
}{\large{}its free most general one-parameter Lagrangian reads}{\large \par}

\begin{eqnarray}
\mathcal{L}(A) & = & \overline{\Psi}_{\mu}(x)\Lambda^{\mu\nu}(A)\Psi_{\nu}(x),\label{eq:Deltafree}
\end{eqnarray}
where

\begin{eqnarray}
\Lambda^{\mu\rho}(A) & = & R_{\hspace{0.2cm}\sigma}^{\mu}(\frac{1+3A}{2})\Lambda^{\sigma\delta}(-\frac{1}{3})R_{\hspace{0.2cm}\delta}^{\rho}(\frac{1+3A}{2}),\label{eq:Kinetic1}\\
\Lambda^{\mu\rho}(-\frac{1}{3}) & = & \left(i\slashed{\partial}-m_{\Delta}\right)g^{\mu\rho}+i\gamma^{\mu}\slashed{\partial}\gamma^{\rho}-i\left(\partial^{\mu}\gamma{}^{\rho}+\gamma^{\mu}\partial^{\rho}\right)+m_{\Delta}\gamma^{\mu}\gamma^{\rho},\nonumber 
\end{eqnarray}
and $R^{\rho\sigma}(a)=g^{\rho\sigma}+a\gamma^{\rho}\gamma^{\sigma}$.
We have that $\Psi_{\mu}\equiv\psi\otimes\xi_{\mu}$, where $\psi$
is a Dirac spinor field and $\xi_{\mu}$ is a Dirac 4-vector \cite{Kirbach2002}.
In this way, the field $\Psi_{\mu}$ will contain a physical spin-
$\frac{3}{2}$ sector and a spurious spin- $\frac{1}{2}$ sector dragged
by construction. The free Lagrangian leads to an equation of motion
\begin{eqnarray}
\Lambda^{\mu\nu}(A)\Psi_{\nu}(x) & = & 0,\label{eq:eqmothion}
\end{eqnarray}
plus certain constraints 
\begin{eqnarray}
\partial_{\mu}\Psi^{\mu} & = & \gamma_{\mu}\Psi^{\mu}=0,\label{eq:constraints}
\end{eqnarray}
independent of $A$ that fix the $\frac{3}{2}$ component eliminating
the redundant $\frac{1}{2}$ contributions. $R^{\rho\sigma}(a)$ changes
the proportion of the spurious $\frac{1}{2}$ component of the Rarita
Schwinger field $\Psi_{\mu}$ but, due to the mentioned constraints,
does not affect the $\frac{3}{2}$ sector. This is the origin of the
family of Lagrangians and $\mathcal{L}(A=-\frac{1}{3})$, where $R^{\rho\sigma}(a=0)=g^{\rho\sigma}$,
was the original form proposed by Rarita-Schwinger \cite{Rarita}.
Also using the properties of $R^{\rho\sigma}$, it should be invariant
under the contact transformation 
\begin{eqnarray}
\Psi^{\nu} & \rightarrow\Psi^{'\nu}= & R_{\mu\nu}(a)\Psi^{\nu},~~A\rightarrow A'=\frac{A-2a}{1+4a},\label{eq:contact}
\end{eqnarray}
$(a\neq-\frac{1}{4},A\neq-\frac{1}{2})$, to avoid a singularity,
and we get $\mathcal{L}(A')=\overline{\Psi}_{\mu}'(x)\Lambda^{\mu\nu}(A')\Psi_{\nu}'(x).$
The invariance of the free Lagrangian under the contact transformations
means that the physical quantities as energy and momentum should be
independent of A. The spin-$\frac{3}{2}$ propagator $G(p,A)^{\beta\nu}$
should satisfy (in momentum space, we replace $i\partial\rightarrow p$),
\[
\Lambda(p_{\Delta},A)^{\beta\mu}G^{\Delta}(p_{\Delta},A)_{\beta\nu}=g^{\mu\nu},
\]
for any value of $A$ and to keep consistence, it should be transformed
as

\begin{eqnarray}
G^{\Delta}(p_{\Delta},A)_{\mu\nu} & = & R^{-1}(\frac{1+3A}{2})_{\mu\alpha}G^{\Delta\alpha\beta}(p_{\Delta},-\frac{1}{3})R^{-1}(\frac{1+3A}{2})_{\beta\nu}\nonumber \\
 & = & R^{-1}(-\onehalf(1+A))_{\alpha}^{\mu}G^{\Delta}(p_{\Delta},-1)^{\alpha\beta}R^{-1}(-\onehalf(1+A))_{\beta}^{\nu}.\label{propagA}
\end{eqnarray}
It can be put in terms of the spin $\frac{3}{2},\frac{1}{2}$ projectors
defined in the appendix A, as (omitting Dirac indexes ) 
\begin{eqnarray}
G_{0}^{\Delta}(p_{\Delta}) & \equiv & G^{\Delta}(p_{\Delta},-\frac{1}{3})=-\left[\frac{\slashed{p}_{\Delta}+m}{p_{\Delta}^{2}-m_{\Delta}^{2}}P^{\frac{3}{2}}+\frac{2}{m_{\Delta}^{2}}(\slashed{p}_{\Delta}+m_{\Delta})P_{11}^{\frac{1}{2}}+\frac{\sqrt{3}}{m_{\Delta}}(P_{12}^{\frac{1}{2}}+P_{21}^{\frac{1}{2}})\right]\label{eq:G(-1/3)}\\
G^{\Delta}\left(p_{\Delta},-1\right) & = & -\left[\frac{\ps_{\Delta}+m}{p_{\Delta}^{2}-m_{\Delta}^{2}}P^{\frac{3}{2}}-\frac{2}{3m_{\Delta}^{2}}(\slashed{p}_{\Delta}+m_{\Delta})P_{22}^{\frac{1}{2}}+\frac{1}{\sqrt{3}m_{\Delta}}(P_{12}^{\frac{1}{2}}+P_{21}^{\frac{1}{2}})\right],\label{eq:G(-1)}
\end{eqnarray}
where we put in terms of $A=-\frac{1}{3},-1$ since will be the cases
to discuss below, or alternatively the developed form 
\begin{eqnarray}
G_{_{\alpha\beta}}^{\Delta}(p_{\Delta},A) & = & -\frac{1}{p_{\Delta}^{2}-m_{_{\Delta}}^{2}}\Bigg\{\left(\slashed{p}_{\Delta}+m_{_{\Delta}}\right)\left[-g_{_{\alpha\beta}}+\frac{1}{3}\gamma_{_{\alpha}}\gamma_{_{\beta}}+\frac{1}{3m_{_{\Delta}}}(\gamma_{_{\alpha}}p_{_{\Delta\beta}}-\gamma_{_{\beta}}p_{_{\Delta\alpha}})+\frac{2}{3m_{_{\Delta}}}p_{_{\Delta\alpha}}p_{_{\Delta\beta}}\right]\nonumber \\
 & - & \frac{2(p_{\Delta}^{2}-m_{_{\Delta}}^{2})b(A)}{3m_{_{\Delta}}^{2}}\left[\gamma_{_{\alpha}}p_{_{\Delta\beta}}-(b(A)-1)\gamma_{_{\beta}}p_{_{\Delta\alpha}}-(\frac{b(A)}{2}\slashed{p}_{\Delta}+(b(A)-1)m_{_{\Delta}})\gamma_{_{\alpha}}\gamma_{_{\beta}}\right]\Bigg\},\label{eq:propagator}
\end{eqnarray}
where $b(A)=\frac{A+1}{2A+1}$. Note that (\ref{eq:propagator}) is
singular at $p^{2}=m_{\Delta}^{2}$, that is when the resonance is
on-shell, but we know that it must be dressed trough a self energy
and thus this singularity it is avoided as we will discuss below.
It is interesting to note that the second A-dependent contribution
in brackets disappears for the $\Delta$ on shell, i.e when the constraints
filter the $\frac{1}{2}$ contribution. Nevertheless, the resonance
appears always off-shell in the presence of interactions, and (\ref{eq:constraints})
do not hold. The amplitudes can be defined uniquely, independent of
$A$, when the interactions with the spin-$\frac{3}{2}$ field are
properly chosen. Consequently, we demand the interaction Lagrangian
for the $\frac{3}{2}$ field coupled to a nucleon ($\psi$) and a
pseudo-scalar meson ($\phi$) or boson ($W$), as usually appears
in a resonance production-decay, be invariant under (\ref{eq:contact}).
The most general interaction Lagrangian satisfying such requirement
is 
\begin{equation}
{\cal {L}}_{int}(A,Z)=g_{int}\bar{\Psi}^{\mu}R(\frac{1}{2}(2Z+(1+4Z)A))_{\mu\nu}F^{\nu}(\psi,\phi,W,...)+h.c.,\label{Lint}
\end{equation}
where $F_{\nu}$ is a function of the fields and its derivatives,
$g_{int}$ is the coupling constant and $Z$ a new arbitrary parameter.
Using the property $R(a)_{\mu\nu}R(b)_{\lambda}^{\nu}=R(a+b+4ab)_{\mu\lambda}$
it is possible to demonstrate 
\begin{equation}
R(\frac{1}{2}(2Z+(1+4Z)A)_{\alpha\beta}=R(\frac{1+3A}{2})_{\alpha\mu}R^{-1}(\frac{1}{2}(1-6Z/(1+4Z))_{\beta}^{\mu}\label{eq:Zparam}
\end{equation}
that would be replaced in Eq.(\ref{Lint}) . Note that the A-dependence
introduced by the propagator (\ref{propagA}) in the $W\psi\rightarrow\phi\psi$
amplitude is canceled by the $R(\frac{1+3A}{2})$ in the vertex generated
from (\ref{Lint}). That is, for any value of $Z$ we get an $A$-independent
amplitude. Then, the value for $Z$ must be chosen for each interaction
and fixed by a criteria independent from contact transformations.

For the strong $\Delta\pi N$ interaction Lagrangian we adopt the
usual chiral invariant one derivative in the pion field

\begin{eqnarray}
\mathcal{L}_{\Delta N\pi}(A=-\frac{1}{3},Z=\frac{1}{2}) & = & \frac{f_{\pi N\Delta}}{m_{\pi}}\bar{\psi}\partial[\bm{\Phi}(x)^{\mu}]^{\dagger}\cdot\bm{T}\Psi_{\mu}+\frac{f_{\pi N\Delta}}{m_{\pi}}\bar{\Psi}_{\mu}\bar{\psi}\partial\bm{\Phi}(x)^{\mu}\cdot\bm{T}^{\dagger}\psi,\label{piNDEltaLag}
\end{eqnarray}
where the choosing in $Z$ will be explained below, and this Lagrangian
enables the definition of the $\Delta\rightarrow\pi N$ vertex

\begin{eqnarray}
V^{\Delta\pi N} & =- & \frac{f_{\pi N\Delta}}{m_{\pi}}k^{\mu}N^{\dagger}\left(\bm{\phi^{*}}\cdot\bm{T}\right)\Delta,\label{piNDeltavertex}
\end{eqnarray}
where we use the prescription $\hat{\Gamma}=i\mathcal{L}$ , $\partial^{\mu}\phi=-ik^{\mu}\phi$
and $i\times propagator$, and a global $i$ in the total amplitudes.

The weak interaction Lagrangian $\hat{\mathcal{L}}_{_{WN\Delta}}$
compatible with the free $\hat{\mathcal{L}}_{_{\Delta}}$ and the
strong interacting Lagrangian $\hat{\mathcal{L}}_{\Delta\pi N}$ that
makes possible also a definition of the weak $WN\Delta$ excitation
vertex, is \cite{Barbero08}(we choose the same $Z$ value) 
\begin{eqnarray*}
\mathcal{L}_{_{WN\Delta}}(A=-\frac{1}{3},Z=\frac{1}{2}) & = & \bar{i\Psi}^{\mu}(x)\hat{{\mathcal{W}}}_{\mu\nu}\sqrt{2}(\bm{T}^{\dagger}\cdot\bm{W}^{\nu}(x)^{\dagger})\psi(x)+\text{h.c.},
\end{eqnarray*}
with a vertex $W^{WN\Delta}=\left(V^{WN\Delta}+A^{WN\Delta}\right)\sqrt{2}\bm{W}^{*}\cdot\bm{T}^{\dagger}$
being the same $V^{WN\Delta}$ vector vertex as in pion-photo($Q^{2}=0)$\cite{Mariano07}
and electroproduction applying CVC

\begin{equation}
V_{\nu\mu}^{WN\Delta}(q,p)=[(G_{_{M}}(Q^{2})-G_{_{E}}(Q^{2}))K_{\nu\mu}^{M}+G_{_{E}}(Q^{2})K_{\nu\mu}^{E}+G_{_{C}}(Q^{2})K_{\nu\mu}^{C}],\label{eq:vectorDelta}
\end{equation}
with $Q^{2}=-q^{2}=-m_{\ell}+2E_{\ell}E_{\nu}(p_{\ell}/E_{\ell}\cos\theta_{\nu\ell})>0$
, being $q=p_{l}-p_{\nu}$, and where 
\begin{align}
K_{\nu\mu}^{M} & =-\frac{3(m_{_{N}}+m_{\Delta})}{2m_{_{N}}(m_{_{N}}+m_{\Delta})^{2}+Q^{2}}\epsilon_{\nu\mu\alpha\beta}\frac{(p+p_{\Delta})^{\alpha}}{2}q^{\beta},\nonumber \\
K_{\nu\mu}^{E} & =\frac{4}{(m_{\Delta}-m_{_{N}})^{2}+Q^{2}}\frac{3(m_{_{N}}+m_{\Delta})}{2m_{_{N}}(m_{_{N}}+m_{\Delta})^{2}+Q^{2}}\epsilon_{\nu\lambda\alpha\beta}\frac{(p+p_{_{\Delta}})^{\alpha}}{2}q^{\beta}\epsilon_{\mu\gamma\delta}^{\lambda}p_{_{\Delta}}^{\gamma}q^{\delta}i\gamma_{_{5}}\nonumber \\
K_{\nu\mu}^{C} & =\frac{2}{(m_{\Delta}-m_{_{N}})^{2}+Q^{2}}\frac{3(m_{_{N}}+m_{1520})}{2m_{_{N}}(m_{_{N}}+m_{\Delta})^{2}+Q^{2}}\left[-q^{2}g_{\alpha\mu}+q_{\alpha}q_{\mu}\right]q_{\nu}\frac{(p+p_{_{\Delta}})^{\alpha}}{2}i\gamma_{_{5}}.\label{eq:KME}
\end{align}
For the FF we adopt 
\begin{equation}
G_{_{i}}(Q^{2})=G_{_{i}}(0)G_{V}(Q^{2}),\label{FFDelta}
\end{equation}
and for the axial contribution we use the model given in Ref.(\cite{Barbero08}),
which is compatible with $V_{\nu\mu}^{WN\Delta}$ (it could be, in
principle, obtained by using $-V_{\nu\mu}^{WN\Delta}\gamma_{_{5}}$)
and reads 
\begin{eqnarray}
A_{\nu\mu}^{WN\Delta}(q,p) & = & -i\Bigg[-D_{_{1}}(Q^{2})g_{_{\nu\mu}}+\frac{D_{_{2}}(Q^{2})}{m_{_{N}}^{2}}(p+p_{_{\Delta}})^{\alpha}(g_{_{\nu\mu}}q_{_{\alpha}}-q_{_{\nu}}g_{_{\alpha\mu}})-\nonumber \\
 & - & \frac{D_{_{3}}(Q^{2})}{m_{_{N}}^{2}}p_{\nu}q_{_{\mu}}+i\frac{D_{4}(Q^{2})}{m_{_{N}}^{2}}\epsilon_{\mu\nu\alpha\beta}(p+p_{_{\Delta}})^{\alpha}q^{\beta}\gamma_{_{5}}\Bigg].\label{eq:axialDelta}
\end{eqnarray}
The $G_{_{i}}(Q^{2})$ and $D_{_{i}}(Q^{2})$ FF will also be described
below .

The bare propagator (\ref{eq:propagator}) being singular at $p_{\Delta}^{2}=m_{\Delta}^{2}$
should be dressed by the inclusion of a self-energy ($\Sigma$) giving
to it a width corresponding to an unstable particle. This self-energy
(where usually only Born interaction terms are considered) could include
the lowest order $\pi N$ one-loop contribution as well as other higher
order $\pi N$ irreducible scattering non-pole contributions consistent
with the $\pi N$ scattering amplitude.

The expression for the dressed propagator $G^{\Delta}{}^{\mu\nu}(p_{\Delta})$
can be obtained by solving the Schwinger-Dyson equation satisfied
by its the inverse

\begin{equation}
[(G^{\Delta}{})_{\mu\nu}]^{-1}(p_{\Delta})=[(G_{0}^{\Delta}{})]_{\mu\nu}^{-1}(p_{\Delta})-\Sigma_{\mu\nu}(p_{\Delta}),\label{dyson}
\end{equation}
where $\Sigma^{\mu\nu}(p)$ denotes the self-energy correction of
$\Delta$ as shown in Fig. 2, and $G_{0}^{\Delta}$ is given in Eq.(\ref{eq:G(-1/3)}).

\begin{figure}
\includegraphics[scale=0.4]{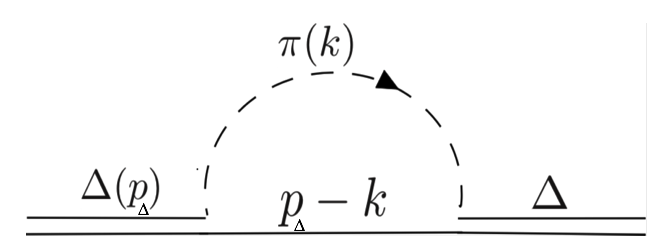}\caption{$\pi N$ loop contribution to the $\Delta$ self energy}
\end{figure}

In the following we will consider only the absorptive (imaginary)
parts of the self-energy correction, i.e. we will assume as in Ref.\cite{Barbero12}
that the parameter $m_{\Delta}$ represents the 'renormalized' mass
of $\Delta$. We place quotation marks as a reminder that the Lagrangian
is not renormalizable; only the absorptive corrections are finite
in this case. Nevertheless, we have analyzed the effect of the real
energy dependent self-energy contribution through dispersive relations
and we found that the effect is small \cite{Barbero15}. If we compute
the one-loop absorptive corrections in Fig.(2) by applying the cutting
rules, we obtain ($g_{int}=\frac{f_{\pi N\Delta}}{m_{\pi}}$)

\begin{equation}
\Sigma_{\text{abs}}^{\mu\nu}(p_{\Delta})=i\frac{g_{int}^{2}}{2(2\pi)^{2}}\int\frac{d^{3}\text{k}}{2\text{k}_{0}}\frac{1}{2\sqrt{s}}\delta\Bigg(\text{k}_{0}+\frac{s+m_{_{\pi}}^{2}-m_{_{N}}^{2}}{2\sqrt{s}}\Bigg)\theta(s-(m_{_{N}}+m_{_{\pi}})^{2})(\slashed{p}_{\Delta}+\slashed{\text{k}}+m_{_{N}})\text{k}^{\mu}\text{k}^{\nu},\label{abs}
\end{equation}
being $s=p_{\Delta}^{2}$ and when developed in terms of the projectors
we can get the corresponding coefficients by solving Eq.(\ref{dyson}),
and the dressed propagator can be finally obtained (for details see
Ref.\cite{Barbero12}). Now, we discuss some approximations commonly
adopted. If neglected terms of $\mathcal{O}(g_{int}^{3})$ and $\mathcal{O}((m_{\Delta}-\sqrt{s})g_{int}^{2})$
in the dressed propagator expression (see Ref.\cite{Barbero15}),
since these terms are expected to very small in the in the resonance
region $(\sqrt{s}\approx m_{\Delta})$, we get again $G_{0}^{\Delta}$
with the replacement 
\begin{eqnarray}
m_{\Delta} & \rightarrow & m_{\Delta}-i\frac{\Gamma_{_{\Delta}}(s)}{2}\nonumber \\
\Gamma_{_{\Delta}}(s) & = & \frac{g_{int}^{2}}{4\pi}\left(\frac{(\sqrt{s}+m_{_{N}})^{2}-m_{_{\pi}}^{2}}{48s^{5/2}}\right))\lambda^{\frac{3}{2}}(s,m_{_{N}}^{2},m_{_{\pi}}^{2}),\label{eq:Gamma_s}\\
\lambda(x,y,z) & = & x^{2}+y^{2}+z^{2}-2xy-2xz-2yz.\nonumber 
\end{eqnarray}

For the sake of completeness, we mention that up to this moment we
have considered only the dressing of the \textgreek{D} propagator.
Nevertheless, analyzing the formal scattering T-matrix calculations
\cite{Mariano07}, one can realize that the $\Delta\pi N$ vertex
should be also dressed by the rescattering. This of course generates
a dependence on $s$ in the vertex, or equivalently an effective coupling
constant $g_{int}(s)$, due to the decay in non-resonant amplitudes
\cite{Mariano07} mediated by the intermediate $\pi N$ propagator.

Now, we consider the formal limit of massless $N$ and $\pi$ in the
loop contribution to $\Sigma$ and in the dressed $\pi N\Delta$ vertex,
this is the so-called complex-mass scheme (CMS) \cite{Amiri92}. It
assumes within this formal limit that the dressing gives a dependence
$g_{int}(s)=\frac{\kappa g_{int}^{0}}{\sqrt{s}}$ ,\cite{Barbero15}
with $g_{int}^{0}$ being the bare $\pi N\Delta$ coupling constant
and $\kappa$ a constant of dimension MeV$^{-1}$ to fit, avoiding
the direct calculation of the momentum integral in the vertex correction.
Thus, we derive from (\ref{eq:Gamma_s}) the following approximated
expression for the width 
\begin{eqnarray}
\Gamma_{\Delta}(s) & = & \left(1-\frac{\sqrt{s}-m_{\Delta}}{m_{\Delta}}\right)\Gamma_{\Delta}^{CMS},\;\Gamma_{\Delta}^{CMS}=\frac{\kappa^{2}(g_{int}^{0})^{2}}{192\pi}m_{\Delta}.\label{eq:CMS}
\end{eqnarray}
In the $s\simeq m_{\Delta}^{2}$region we have a constant width $\Gamma_{\Delta}(s)\approx\Gamma_{\Delta}^{CMS}$,
where now $\Gamma_{\Delta}^{CMS}$ is fitted in place of $\kappa$
together $g_{int}$ and $m_{\Delta}$ to reproduce $\pi^{+}p$ scattering
\cite{Mariano01}. Note that we have the isospin coefficient in the
previous equation equal to 1 for the $\Delta^{++}\rightarrow\pi^{+}p\rightarrow\Delta^{++}$
loop as was shown in the Appendix A. Another approach commonly used,
is to fix $\sqrt{s}\approx m_{\Delta}$ in (\ref{eq:Gamma_s}) and
to use the experimental values for $m_{\Delta}$ and $\Gamma_{\Delta}$
times the branching ratio for the decay into $\pi N$, and get $g_{int}$.
We will refer to this as constant mass-width approach (CMW). We will
use both the CMS and CMW depending on the considered resonance.

\subsubsection{$N^{*}(1520)$ resonance}

This $IJ^{\pi}=\frac{1}{2},\frac{3}{2}^{-}$ negative parity resonance
has three quark orbital momentum and spin $L=1,S=\frac{1}{2}$. The
propagator is (\ref{eq:G(-1/3)}) but changing $m_{\Delta}\rightarrow m_{1520}$,
where we will use the notation $N^{*}(1520)\equiv1520$.\textsf{\large{}{}
}\textcolor{black}{The rescattering in the propagator will be introduced
making $m_{_{1520}}\rightarrow m_{_{1520}}-i\frac{\Gamma_{1520}}{2}$
and since in the second resonance region $W_{\pi N'}=\sqrt{(p_{N'}+k)^{2}}\lesssim1600{\text{MeV}}\sim m_{1520}+\Gamma_{1520}$
we can adopt the the CMW with $m_{1520}=1529$ MeV and $\Gamma_{1520}=\Gamma_{1520}^{N\pi}+\Gamma_{1520}^{\Delta\pi}=115$
MeV \cite{PTEP20}.The strong Lagrangian is given by(\cite{Leitner09}):
\begin{eqnarray}
\mathcal{L}_{1520\pi N} & = & =\frac{f_{_{1520\pi N}}}{m_{\pi}}\bar{\Psi}_{\mu}\gamma_{_{5}}\partial^{\mu}\bm{\Phi}_{\pi}(x)\cdot\bm{\tau}{\Psi}-\frac{f_{_{1520\pi N}}}{m_{\pi}}\bar{\Psi}\partial^{\mu}\bm{\Phi}_{\pi}^{\dagger}(x)\cdot\bm{\tau}\gamma_{5}\Psi_{\mu},\label{eq:1520Npi}
\end{eqnarray}
where $\Psi_{\mu}$ is a Rarita - Schwinger field for the spin-$\frac{3}{2}$
but isospin $\frac{1}{2}$. Note that is the same Lagrangian (\ref{piNDEltaLag})
but with $\gamma_{5}$ inserted and changing $\bm{T}\rightarrow\bm{\tau}$
. From this Lagrangian we derive the $N^{*}(1520)\rightarrow\pi N$
vertex decay}

\begin{eqnarray*}
V^{^{1520N\pi}} & = & \frac{f_{_{1520\pi N}}}{m_{\pi}}{\text{k}}_{\mu}\gamma_{_{5}}\left(\bm{\Phi}_{\pi}^{^{*}}\cdot\bm{\tau}\right),
\end{eqnarray*}
and in Eq.(\ref{abs}) we have a minus sign in the $\ps+\ks$ term
due the $\gamma_{5}$ in the vertex and changing the isospin coefficients
to 3 since we have now $\pi^{0}p$ and $\pi^{+}n$ intermediate loop
states (see Appendix A), we can get the relation

\begin{eqnarray*}
\Sigma_{\text{abs}}^{\mu\nu}(p)(1520) & = & -\Sigma_{\text{abs}}^{\mu\nu}(p,-m_{\pi,-m_{N)}})(\Delta)
\end{eqnarray*}
this leads to ($g_{int}=\frac{f_{_{1520\pi N}}}{m_{\pi}}$)

\begin{eqnarray}
\Gamma_{_{1520}}(s) & = & \frac{3g_{int}^{2}}{4\pi}\left(\frac{(\sqrt{s}-m_{_{N}})^{2}-m_{_{\pi}}^{2}}{48s^{5/2}}\right)\lambda^{\frac{3}{2}}(s,m_{_{N}}^{2},m_{_{\pi}}^{2})\nonumber \\
 & = & \frac{3g_{int}^{2}}{12\pi}\left(\frac{\frac{(\sqrt{s}-m_{_{N}})^{2}-m_{_{\pi}}^{2}}{2\sqrt{s}}}{\sqrt{s}}\right){\rm q}_{CM}^{3},\;{\rm q}_{CM}=\frac{\lambda^{\frac{3}{2}}(s,m_{_{N}}^{2},m_{_{\pi}}^{2})}{2\sqrt{s}},\label{Gamma1520}
\end{eqnarray}
that within the CMW the approximation $\sqrt{s}\approx m_{1520}$
it is done and we get the expression used in Ref.(\cite{Leitner09}),
where $\Gamma_{1520}$ should be weighted by the corresponding $\pi N$
branching ratio decay.

Usually the vector vertex FF for this resonance are expressed in the
so called parity conserving parametrization \cite{Lakakulich06,Leitner09},
nevertheless we want for consistence to express them in the same Sachs
parametrization as the other present spin- $\frac{3}{2}$ resonance
that is the $\Delta$ . Then, we will assume similar vertex structure
than for $\Delta$ in (\ref{eq:vectorDelta}) times $\gamma_{5}$
(for the changing in parity), then transform to parity conserving
parametrization, compare with Ref.(\cite{Leitner09}) and fix our
parameters. We get the axial vertex multiplying by $\gamma_{5}$ the
$\Delta$ one (\ref{eq:axialDelta}). We get $W^{WN1520}=\left(V^{WN1520}+A^{WN1520}\right)\frac{\sqrt{2}}{2}\left(\bm{W}^{*}\cdot\boldsymbol{\tau}\right)$
with

\begin{eqnarray}
V_{\nu\mu}^{WN1520}(q,p) & = & \left[(G_{M}(Q^{2})-G_{E}(Q^{2}))K_{\nu\mu}^{M}+G_{E}(Q^{2})K_{\nu\mu}^{E}+G_{C}(Q^{2})K_{\nu\mu}^{C}\right]\gamma_{5},\label{eq:vector1520}\\
A_{\nu\mu}^{WN1520}(q,p) & = & i\left[D_{1}(Q^{2})g_{\nu\mu}-\frac{D_{2}(Q^{2})}{m_{_{N}}^{2}}(p+p_{1520})^{\alpha}(g_{\mu\nu}q_{\alpha}-q_{\nu}g_{\mu\alpha})+\frac{D_{3}(Q^{2})}{m_{N}^{2}}q_{\mu}p_{\nu}\right]\gamma_{5},\label{eq:axial1520}
\end{eqnarray}
where $K^{i}(q,p)$, $G_{i}(Q^{2})$ and $D_{i}(Q^{2})$ are the same
that in Eqs.(\ref{eq:vectorDelta}) and (\ref{eq:axialDelta}) but
changing $m_{\Delta}\rightarrow m_{1520},p_{\Delta}\rightarrow p_{1520}$,
and the values $G_{i}(0),D_{i}(0)$. Note that we have an additional
$\frac{1}{2}$ factor coming from the charge operator $\hat{q}=\frac{1+\tau_{3}}{2}$
present in the isospin $\frac{1}{2}$ electromagnetic vertexes but
not in the $\frac{3}{2}$ case where we have $T_{3}^{\dagger}$ transition
operators.

Now, $V_{\nu\mu}$(we omit super indexes) can be expressed in the
so-called ``normal parity\textquotedbl{} (NP) decomposition making
use of non-trivial relation

\begin{eqnarray*}
-i\epsilon_{\alpha\beta\mu\nu}a^{\mu}b^{\nu}\gamma_{_{5}} & = & (\slashed{a}\slashed{b}-a\cdot b)i\sigma_{\alpha\beta}+\slashed{b}(\gamma_{\alpha}a_{\beta}-\gamma_{\beta}a_{\alpha})-\slashed{a}(\gamma_{\alpha}b_{\beta}-\gamma_{\beta}b_{\alpha})\\
 & - & \slashed{a}(\gamma_{\alpha}b_{\beta}-\gamma_{\beta}b_{\alpha})+(a_{\alpha}b_{\beta}-a_{\beta}b_{\alpha}),
\end{eqnarray*}
and some on-shell considerations on the resonance, we getting a simplified
version of $V_{\nu\mu}$ \cite{Barbero15}

\begin{eqnarray}
V_{\nu\mu}(q,p) & = & i\Bigg\{-(G_{M}(Q^{2})-G_{E}(Q^{2}))m_{1520}H_{3\nu\mu}+\Bigg[G_{M}(Q^{2})-G_{E}(Q^{2})+2\frac{2G_{E}(Q^{2})(q\cdot p_{1520})-G_{C}(Q^{2})Q^{2}}{(m_{1520}-m_{_{N}})^{2}+Q^{2}}\Bigg]H_{4\nu\mu}\nonumber \\
 & - & \Bigg[2\frac{2G_{E}(Q^{2})m_{1520}^{2}+(p_{1520}\cdot q)G_{C}(Q^{2})}{(m_{1520}-m_{_{N}})^{2}+Q^{2}}\Bigg]H_{6\nu\mu}\Bigg\}\frac{3(m_{_{N}}+m_{1520})}{2m_{_{N}}(m_{_{N}}+m_{1520})^{2}+Q^{2}},\label{eq:SachParityV}
\end{eqnarray}
where

\begin{eqnarray*}
H_{3}^{\nu\mu}(p,q) & = & g^{\nu\mu}\qs-q^{\nu}\gamma^{\mu},\\
H_{4}^{\nu\mu}(p,q) & = & g^{\nu\mu}q\cdot p_{1520}-q^{\nu}p_{1520}^{\mu},\\
H_{5}^{\nu\mu}(p,q) & = & g^{\nu\mu}q\cdot p-q^{\nu}p^{\nu},\\
H_{6}^{\nu\mu}(p,q) & = & g^{\nu\mu}q^{2}-q^{\nu}p^{\nu}.
\end{eqnarray*}

Note that the $H_{5}^{\nu\mu}$ tensor does not contribute to Eq.(\ref{eq:SachParityV}),
but it appears in the general Parity-Conserving expression. The Eq.(\ref{eq:vector1520})
are independent of taking $p_{1520}=p\mp q$ or $p=p_{1520}\pm q$,
(see Eq.(\ref{eq:KME})) $+$ sign corresponds to the pole contribution
and $-$ sign to the cross term. Thus, the Eq(\ref{eq:SachParityV})
is valid in both cases, but the specific value of $q\cdot p_{1520}$
depends on the particular contribution to the amplitudes $\Bigg(q\cdot p_{1520}=\pm\frac{m_{_{N}}^{2}+Q^{2}-m_{1520}^{2}}{2}\Bigg)$.
If we set on the $N^{*}(1520)$-pole contribution and replace $p=p_{1520}+q$
we can write Eq. (\ref{eq:SachParityV}) as usually in the parity
conserving form 
\begin{eqnarray}
V_{\nu\mu}(p,q) & = & i\Gamma_{\nu\mu}^{V}(p,q),\nonumber \\
\Gamma_{\nu\mu}^{V}(p_{_{\text{D}_{13}}},q) & = & \Bigg[-\frac{C_{3}^{V}(Q^{2})}{m_{_{N}}}H_{3\nu\mu}-\frac{C_{4}^{V}(Q^{2})}{m_{_{N}}^{2}}H_{4\nu\mu}-\frac{C_{5}^{V}(Q^{2})}{m_{_{N}}^{2}}H_{5\nu\mu}+\frac{C_{6}^{V}(Q^{2})}{m_{_{N}}^{2}}H_{6\nu\mu}\Bigg],\label{eq:Vparitycons}
\end{eqnarray}
where we have the corresponding FF:

\begin{align}
C_{3}^{V}(Q^{2}) & =\frac{m_{1520}}{m_{_{N}}}R_{M}\Bigg[G_{M}(0)-G_{E}(0)\Bigg]F^{V}(Q^{2})\nonumber \\
C_{4}^{V}(Q^{2}) & =-R_{M}\Bigg[G_{M}(0)-\frac{3m_{1520}}{m_{1520}-m_{_{N}}}G_{E}(0)\Bigg]F^{V}(Q^{2})\nonumber \\
C_{5}^{V}(Q^{2}) & =0\nonumber \\
C_{6}^{V}(Q^{2}) & =-R_{M}\frac{2m_{1520}}{m_{1520}-m_{_{N}}}G_{E}(0)F^{V}(Q^{2}),\label{eq:CVGME}
\end{align}
being $R_{M}=\frac{{3}}{{2}}\frac{m_{_{N}}}{m_{1520}+m_{_{N}}}$ and
$F^{V}(Q^{2})=\bigg(1+\frac{Q^{2}}{(m_{_{N}}+m_{1520})^{2}}\bigg)^{-1}G^{V}(Q^{2})$.
Note that $\Gamma_{\nu\mu}^{V}(p,q)$ coincides with Eqs.(30) and
(31) in Ref. (\cite{Leitner09}) making $q\rightarrow-q$ and that
now taking the values for $C_{i}^{V}(0)$ from that reference we can
get $G_{M,E}(0)$ for the $N^{*}(1520)$ resonance. Rearranging Eq.(\ref{eq:axial1520})we
get for the pole case

\begin{eqnarray*}
A_{\nu\mu}(p,q) & = & \sqrt{2}i\Bigg[\Bigg(D_{1}(Q^{2})+\frac{D_{2}(Q^{2})Q^{2}}{m_{_{N}}^{2}}\Bigg)g_{\nu\mu}-\frac{2D_{2}}{m_{_{N}}^{2}}H_{4\nu\mu}+\frac{D_{3}(Q^{2})+D_{2}(Q^{2})}{m_{_{N}}^{2}}q_{\nu}q_{\mu}\Bigg]\gamma_{5}=i\Gamma_{\mu\nu}^{A},
\end{eqnarray*}

\begin{align}
\Gamma_{\mu\nu}^{A} & =\left[C_{5}^{A}g_{\nu\mu}-\frac{C_{4}^{A}}{m_{_{N}}^{2}}g^{\mu\nu}H_{4\nu\mu}+\frac{C_{6}^{A}}{m_{_{N}}^{2}}H_{6\mu\nu}\right]\gamma_{5}.\label{eq:axialparitycons}
\end{align}
where as before $D_{4}(Q^{2})=0$. Note that this last coincides with
Eq.(32) from Ref.\cite{Leitner09} making $q\rightarrow-q$. By comparison
we get

\begin{eqnarray}
D_{1}+D_{2}\frac{Q^{2}}{m_{_{N}}^{2}} & = & C_{5}^{A},\nonumber \\
-\frac{2D_{2}}{m_{_{N}}^{2}} & = & -\frac{C_{4}^{A}}{m_{_{N}}^{2}},\nonumber \\
0 & = & C_{3}^{A}\nonumber \\
\frac{D_{3}+D_{2}}{m_{_{N}}^{2}} & = & \frac{C_{A}^{6}}{m_{_{N}}^{2}}.\label{eq:DGME}
\end{eqnarray}
From Eqs.(\ref{eq:DGME})one can get from Ref.(\cite{Leitner09})
the $D_{i}$ values for the axial resonance $N^{*}(1520)$vertex.

\subsection{Spin $\frac{1}{2}$ resonances}

For the considered resonances of spin $\frac{1}{2}$ that has three
quark orbital momentum and spin $L=0,1,S=\frac{1}{2}$, the parametrization
of the hadronic vertex is simpler than for spin- $\frac{3}{2}$ ones
and is similar to the parametrization for the $\nu N\rightarrow N'$
vertex depending on the parity. We will include the $L=0$, $IJ^{\pi}=\frac{1}{2},\frac{1}{2}^{+}$
$N^{*}(1440)$ resonance and the $L=1$, $IJ^{\pi}=\frac{1}{2},\frac{1}{2}^{-}N^{*}(1535)$
one. The propagator of these resonances looks like the nucleon one
but with the replacement $m_{R}\rightarrow m_{R}-i\frac{\Gamma_{R}}{2}$
to introduce the width, and $\Gamma_{R}$ will be considered constant(CMW)
since the second resonance region extends to $1600$ MeV and close
to the centroids. We get 
\begin{eqnarray}
S_{R}(p) & = & \frac{\slashed{p}+m_{R}}{p^{2}-m_{R}^{2}+i\Gamma_{R}m_{R}},\label{eq:propRspin_1/2}
\end{eqnarray}
where $\Gamma_{R}$ should be weighted by the corresponding $\pi N$
branching ratio decay. The $RN\pi$ strong coupling is described by
the Lagrangian \cite{Leitner09}: 
\begin{equation}
\mathcal{L}_{RN\pi}=\frac{f_{{R\pi N}}}{m_{\pi}}\left(\bar{\Psi}_{R}\gamma^{\mu}\Lambda\bm{\tau}\Psi_{N}\right)\cdot\partial_{\mu}\bm{\Phi}_{\pi}(x)+\frac{f_{R\pi N}}{m_{\pi}}\partial_{\mu}\bm{\Phi}_{\pi}^{\dagger}(x)\cdot\left(\bar{\Psi}_{N}\gamma^{\mu}\Lambda\bm{\tau}\Psi_{P_{11}}\right),\label{eq:piNR_1/2}
\end{equation}
where $\Lambda=\gamma_{5},I$ for positive or negative parity. Note
that in the $N^{*}(1440)$ case this Lagrangian is similar to the
$\mathcal{L}_{NN\pi}$ . From the Lagrangian (\ref{eq:piNR_1/2}),
we can deduce the $R\pi N$ decay vertex 
\begin{equation}
V^{R\pi N}=\pi\frac{f_{_{R\pi N}}}{m_{\pi}}\Lambda\slashed{\text{k}}(\bm{\Phi}_{\pi}^{*}\cdot\bm{\tau}).\label{eq:R_1/2piN}
\end{equation}
For the $W\rightarrow NR$ vertex as we have an outgoing boson, make
$q\rightarrow-q$ in the hadronic vertex of Ref.(\cite{Lakakulich06}),
and the vertex can be written as $W^{WNR\lambda}\times\sqrt{2}\bm{\tau}\cdot\bm{W}^{*}$
with

\begin{align}
W^{WNR\lambda} & =-i\frac{1}{2}\Bigg[\frac{g_{1V}}{(m_{_{R}}+m_{_{N}})^{2}}(Q^{2}\gamma^{\lambda}+\slashed{q}q^{\lambda})\gamma_{5}-\frac{g_{2V}}{(m_{_{R}}+m_{_{N}})}i\sigma^{\lambda\nu}q_{\nu}\gamma_{5}-g_{1A}\gamma^{\lambda}+\frac{g_{3A}}{m_{_{N}}}q^{\lambda}\Bigg]\Lambda\nonumber \\
\Lambda & =\gamma_{5},I,\mbox{for parity }\pi=\pm1,\label{WNP11}
\end{align}
where we note that Eq.(\ref{WNP11}) is the same as in Ref.(\cite{Leitner09})
making $q\rightarrow-q$ but changing $m_{R}+m_{N}\rightarrow2m_{N}$
and $g_{1V}$, $g_{2V}$, $g_{1A}$, $g_{3A}$ $\rightarrow$ $\mathcal{F}_{1}$,
$\mathcal{F}_{2}$, $-\mathcal{F}_{A}$, $-\mathcal{F}_{P}$. We note
the similarity of (\ref{WNP11}) with the same for nucleons which
will be shown in the calculations of the background contributions
in the next section.

The term with $g_{3A}$ is called pion-pole term and gives de contribution
where the W boson decays in a pion which then interacts with the nucleon.
This can be obtained replacing the axial contribution $A^{\lambda}$
by $A^{\lambda}+q^{\mu}q.A/(Q^{2}+m_{\pi}^{2})$ (see $g_{3A}(Q^{2})$
below). Then, one assumes that the resonance is on shell and evaluates
$\bar{u}_{R}(\slashed{q}=\ps-\ps_{R})\Lambda u=\mp\bar{u}_{R}\Lambda(m_{R}\pm m_{N})u$.
The FF for the $W\rightarrow NN^{*}$(1440) vertex are obtained from
the connection between electromagnetic resonance production and the
helicity amplitudes.The helicity amplitudes describe the nucleon-resonances
transition depending on the polarization of the incoming photon and
the spins of the baryons \cite{Lakakulich06}. For non-zero $Q^{2}$,
data on helicity amplitudes for the $N^{*}$ resonance are available
only for the proton \cite{Lakakulich06}, where it is assumed that
the isovector contribution on the neutrino production is given as
$g_{i}^{V}=-2g_{i}^{p}$. The PCAC hypothesis allows us to relate
the two form factors and fix their axial values at $Q^{2}=0$(\cite{Lakakulich06}),
we get

\begin{eqnarray}
g_{1V}(Q^{2}) & = & \frac{g_{1V}(0)\left(b_{1V}ln(1+\frac{Q^{2}}{GeV^{2}})+\frac{1-\pi}{2}\right)}{(1+Q^{2}/M_{V}^{2})^{2}(1+Q^{2}/a_{1V}M_{V}^{2})},\;g_{2V}(Q^{2})=\frac{g_{2V}(0)\left(b_{2V}ln(1+\frac{Q^{2}}{GeV^{2}})-\frac{1+\pi}{2}\right)}{(1+Q^{2}/M_{V}^{2})^{2}}\nonumber \\
g_{1A}(Q^{2}) & = & \frac{g_{1A}(0)}{(1+Q^{2}/M_{A}^{2})^{2}(1+Q^{2}/3M_{A}^{2})},\;g_{3A}(Q^{2})=\frac{g_{1A}(0)(m_{R}+\pi m_{_{N}})}{Q^{2}+m_{\pi}^{2}}m_{N}.\label{eq:gVA}
\end{eqnarray}
The coupling $g=\frac{f_{R\pi N}}{m_{\pi}}$ can be obtained of the
partial decay width ($R\rightarrow\pi N$) according to (\cite{Leitner09})

\begin{eqnarray}
\Gamma_{R\rightarrow\pi N} & = & \frac{3}{4\pi}g^{2}(m_{R}+\pi m_{N})^{2}\left(\frac{\frac{(\sqrt{s}-\pi m_{_{N}})^{2}-m_{_{\pi}}^{2}}{2\sqrt{s}}}{\sqrt{s}}\right){\rm q}_{CM},\;{\rm q}_{CM}=\frac{\lambda^{\frac{1}{2}}(s,m_{_{N}}^{2},m_{_{\pi}}^{2})}{2\sqrt{s}},\label{width1/2}
\end{eqnarray}
using the CMW approach mentioned above.

\section{Background and resonance amplitudes}

Now we built the different components of $\mathcal{O}_{B}$ and $\mathcal{O}_{R}$
from the Lagrangians shown in the Appendix B and those described in
the previous section. We get

\begin{eqnarray*}
O_{\text{B}}^{\lambda}(p,p',q) & = & O_{\text{BN}}^{\lambda}(p,p',q)+O_{\text{BR}}^{\lambda}(p,p',q)
\end{eqnarray*}

\begin{eqnarray}
O_{\text{BN}}^{\lambda}(p,p',q) & = & -i\frac{1}{2}\Bigg[F_{1}^{V}(Q^{2})\gamma^{\lambda}-i\frac{F_{2}^{V}(Q^{2})}{2m_{_{N}}}\sigma^{\lambda\nu}q_{\nu}-F^{A}(Q^{2})\gamma^{\lambda}\gamma_{_{5}}\Bigg]i\frac{\ps'+\qs+m_{_{N}}}{(p'+q)^{2}-m_{_{N}}}\nonumber \\
 & \times & \frac{g_{_{\pi NN}}}{2m_{_{N}}}\gamma_{_{5}}(\ps-\ps'-\qs)\sqrt{2}\mathcal{T}_{a}(m_{t},m_{t'})\nonumber \\
 & + & \frac{g_{_{\pi NN}}}{2m_{_{N}}}\gamma_{_{5}}(\ps-\ps'-\qs)i\frac{\ps-\qs+m_{_{N}}}{(p-q)^{2}-m_{_{N}}^{2}}\left(-i\frac{1}{2}\right)\Bigg[F_{1}^{V}(Q^{2})\gamma^{\lambda}-i\frac{F_{2}^{V}(Q^{2})}{2m_{_{N}}}\sigma^{\lambda\nu}q_{\nu}-F^{A}(Q^{2})\gamma^{\lambda}\gamma_{_{5}}\Bigg]\nonumber \\
 & \times & \sqrt{2}\mathcal{T}_{b}(m_{t},m_{t'})\nonumber \\
 & - & \frac{i}{(p-p')^{2}-m_{\pi}^{2}}iF_{1}^{V}(Q^{2})(2p-2p'-q)^{\lambda}\times\frac{g_{_{\pi NN}}}{2m_{_{N}}}\gamma_{_{5}}(\ps-\ps')\sqrt{2}\mathcal{T}_{c}(m_{t},m_{t'})\nonumber \\
 & + & \frac{g_{_{\pi NN}}}{2m_{_{N}}}F_{1}^{V}(Q^{2})\gamma_{_{5}}\gamma^{\lambda}\sqrt{2}\mathcal{T}_{d}(m_{t},m_{t'})\nonumber \\
 & + & i\frac{g_{_{\omega\pi V}}}{m_{\pi}}F_{1}^{V}(Q^{2})\epsilon^{\lambda\alpha\beta\delta}q_{_{\alpha}}(p-p')_{_{\beta}}i\frac{-g_{_{\delta\epsilon}}}{(p-p')^{2}-m_{\omega}^{2}}(-i)\frac{g_{_{\omega NN}}}{2}\Bigg[\gamma^{\epsilon}-i\frac{\kappa_{\omega}}{2m_{_{N}}}\sigma^{\epsilon\kappa}(p-p')_{\kappa}\Bigg]\nonumber \\
 & \times & \sqrt{2}\mathcal{T}_{e}(m_{t},m_{t'})\nonumber \\
 & + & f_{\rho\pi A}F^{A}(Q^{2})i\frac{-g^{\lambda\mu}}{(p-p')^{2}-m_{\rho}^{2}}(-i)\frac{g_{_{\rho NN}}}{2}\Bigg[\gamma_{\mu}-i\frac{\kappa_{\rho}}{2m_{_{N}}}\sigma_{\mu\kappa}(p-p')^{\kappa}\Bigg]\sqrt{2}\mathcal{T}_{f}(m_{t},m_{t'}),\label{eq:OBN}
\end{eqnarray}
\vspace{-0.5cm}
 
\begin{eqnarray}
O_{\text{BR}}^{\lambda}(p,p',q) & =- & i\frac{1}{2}\Bigg[\frac{g_{1V}^{1440}}{(m_{_{1440}}+m_{_{N}})^{2}}(Q^{2}\gamma^{\lambda}+\qs q^{\lambda})-\frac{g_{2V}^{1440}}{(m_{_{1440}}+m_{_{N}})}i\sigma^{\lambda\nu}q_{\nu}-g_{1A}^{1440}\gamma^{\lambda}\gamma_{5}+\frac{g_{3A}^{1440}}{m_{_{N}}}q^{\lambda}\gamma_{5}\Bigg]\nonumber \\
 & \times & i\frac{\ps'+\qs+m_{R}}{(p'+q)^{2}-m_{1440}^{2}+i\Gamma_{1440}m_{1440}}(-)\frac{f_{_{1440\pi N}}}{m_{\pi}}\gamma_{5}(\ps-\ps'-\qs)\sqrt{2}\mathcal{T}_{g}^{1440}(m_{t},m_{t'})\nonumber \\
 & - & i\frac{1}{2}\gamma_{5}\Bigg[\frac{g_{1V}^{1535}}{(m_{_{1535}}+m_{_{N}})^{2}}(Q^{2}\gamma^{\lambda}+\qs q^{\lambda})-\frac{g_{2V}^{1535}}{(m_{_{1535}}+m_{_{N}})}i\sigma^{\lambda\nu}q_{\nu}-g_{1A}^{1535}\gamma^{\lambda}\gamma_{5}+\frac{g_{3A}^{1535}}{m_{_{N}}}q^{\lambda}\gamma_{5}\Bigg]\nonumber \\
 & \times & i\frac{\ps'+\qs+m_{1535}}{(p'+q)^{2}-m_{1535}^{2}+i\Gamma_{1535}m_{1535}}(-)\frac{f_{_{1535\pi N}}}{m_{\pi}}(\ps-\ps'-\qs)\sqrt{2}\mathcal{T}_{g}^{1535}(m_{t},m_{t'})\nonumber \\
 & + & (-\overline{){W_{\lambda\alpha}^{WN\Delta}(p,p',-q)}}iG_{\Delta}^{\alpha\beta}(p'+q)(-)\frac{f_{_{\pi N\Delta}}}{m_{\pi}}(p-p'-q)_{\beta}\sqrt{2}\mathcal{T}_{g}^{\Delta}(m_{t},m_{t'})\nonumber \\
 & + & (-\overline{){\frac{1}{2}W_{\lambda\alpha}^{WN1520}(p,p',-q)}}iG_{1520}^{\alpha\beta}(p'+q)(-)\frac{f_{_{\pi N1520}}}{m_{\pi}}\gamma_{5}(p-p'-q)_{\beta}\sqrt{2}\mathcal{T}_{g}^{1520}(m_{t},m_{t'}),\label{eq:OBR}
\end{eqnarray}
where the background contributions were splitted in those coming from
the nucleon contributions and those coming from the resonances one.
Here $\mathcal{T}(m_{t},m_{t'})$ are isospin factors calculated between
the initial an final nucleon with isospin $m_{t},m_{t}'$ projections
respectively for each amplitude contribution. Note that the $\frac{1}{2}$
factor in the weak vertex of the isospin $\frac{1}{2}$ resonances
comes from the isovector part of the charge operator $\frac{\tau_{3}}{2}$
dragged from the CVC hypothesis. The corresponding pole contributions
coming from the resonances are

\begin{eqnarray}
\mathcal{O}_{R}^{\lambda} & = & \frac{f_{_{1440\pi N}}}{m_{\pi}}\gamma_{5}(\ps-\ps'-\qs)i\frac{\ps-\qs+m_{R}}{(p-q)^{2}-m_{1440}^{2}+i\Gamma_{1440}m_{1440}}\nonumber \\
 & \times & (-i)\frac{1}{2}\Bigg[\frac{g_{1V}^{1440}}{(m_{_{1440}}+m_{_{N}})^{2}}(Q^{2}\gamma^{\lambda}+\qs q^{\lambda})-\frac{g_{2V}^{1440}}{(m_{_{1440}}+m_{_{N}})}i\sigma^{\lambda\nu}q_{\nu}-g_{1A}^{1440}\gamma^{\lambda}\gamma_{5}+\frac{g_{3A}^{1440}}{m_{_{N}}}q^{\lambda}\gamma_{5}\Bigg]\sqrt{2}\mathcal{T}_{h}^{1440}(m_{t},m_{t'})\nonumber \\
 & + & (-)\frac{f_{_{1535\pi N}}}{m_{\pi}}(\ps-\ps'-\qs)i\frac{\ps-\qs+m_{1535}}{(p-q)^{2}-m_{1535}^{2}+i\Gamma_{1535}m_{1535}}\nonumber \\
 & \times & (-i)\frac{1}{2}\Bigg[\frac{g_{1V}^{1535}}{(m_{_{1535}}+m_{_{N}})^{2}}(Q^{2}\gamma^{\lambda}+\qs q^{\lambda})-\frac{g_{2V}^{1535}}{(m_{_{1535}}+m_{_{N}})}i\sigma^{\lambda\nu}q_{\nu}-g_{1A}^{1535}\gamma^{\lambda}\gamma_{5}+\frac{g_{3A}^{1535}}{m_{_{N}}}q^{\lambda}\gamma_{5}\Bigg]\gamma_{5}\mathcal{T}_{h}^{1335}(m_{t},m_{t'})\nonumber \\
 & + & (-)\frac{f_{_{\pi N\Delta}}}{m_{\pi}}(p-p'-q)_{\alpha}iG_{\Delta}^{\alpha\beta}(p-q)W_{\beta\lambda}^{WN\Delta}(p,p',q)\sqrt{2}\mathcal{T}_{h}^{\Delta}(m_{t},m_{t'})\nonumber \\
 & + & (-)\frac{f_{_{\pi N1520}}}{m_{\pi}}\gamma_{5}(p-p'-q)_{\alpha}iG_{1520}^{\alpha\beta}(p-q)W_{\beta\lambda}^{WN1520}(p,p',q)\sqrt{2}\mathcal{T}_{h}^{1520}(m_{t},m_{t'}).\label{eq:OR}
\end{eqnarray}

Here we show the isospin coefficients calculated with the ingredients
of appendix A

\begin{align}
\mathcal{T}_{a}(m_{t},m_{t'}) & =\mathcal{T}_{g}^{1440,1535,1520}(m_{t},m_{t'})=\chi^{\dagger}(m_{t'})({\bm{\tau}\cdot{\bf W^{*}}})({\bm{\tau}\cdot\bm{\Phi}_{\pi}^{*}})\chi(m_{t})=-2,0,-\sqrt{2}\nonumber \\
\mathcal{T}_{b}(m_{t},m_{t'}) & =\mathcal{T}_{h}^{1440,1535,1520}(m_{t},m_{t'})=\chi^{\dagger}(m_{t'})({\bm{\tau}\cdot\bm{\Phi}_{\pi}^{*}})({\bm{\tau}\cdot{\bf W^{*}}})\chi(m_{t})=0,-2,\sqrt{2}\nonumber \\
\mathcal{T}_{c}(m_{t},m_{t'}) & =-i\chi^{\dagger}(m_{t'})[({\bm{\Phi}_{\pi}^{*}\times\bm{\Phi}_{\pi'}})\cdot{\bf W^{*}}]({\bm{\tau}\cdot}\bm{\Phi}_{\pi'}^{*})\chi(m_{t'})=1,-1,\sqrt{2}\nonumber \\
\mathcal{T}_{d}(m_{t},m_{t'}) & =\chi^{\dagger}(m_{t'})[(\bm{\Phi}_{\pi}^{*}\times\bm{\tau})\cdot{\bf W^{*}}]\chi(m_{t})=-1,1,-\sqrt{2}\nonumber \\
\mathcal{T}_{e}(m_{t},m_{t'}) & =\chi^{\dagger}(m_{t'})(\bm{\Phi}_{\pi}^{*}\cdot{\bf W^{*}})\chi(m_{t})=-1,-1,0\nonumber \\
\mathcal{T}_{f}(m_{t},m_{t'}) & =i\chi^{\dagger}(m_{t'})[(\bm{\Phi}_{\pi}^{*}\times\bm{\rho})\cdot{\bf W^{*}}](\bm{\tau}\cdot\bm{\rho}^{*})\chi(m_{t})=-1,1,-\sqrt{2}\nonumber \\
\mathcal{T}_{g}^{\Delta}(m_{t},m_{t'}) & =\chi^{\dagger}(m_{t'})({\bf T\cdot W^{*}})({\bf T^{\dagger}}\cdot\bm{\Phi}_{\pi}^{*})\chi(m_{t})=-1/3,\sqrt{2}/3,-1\nonumber \\
\mathcal{T}_{h}^{\Delta}(m_{t},m_{t'}) & =\chi^{\dagger}(m_{t'})({\bf T}\cdot\bm{\Phi}_{\pi}^{*})({\bf T^{\dagger}\cdot W^{*}})\chi(m_{t})=-1,-\sqrt{2}/3,-1/3.\label{eq:isospincoef}
\end{align}

\section{Form factors and Results}

In this work we analyze as a first step the total cross section for
the charged current (CC) modes of the six processes 
\begin{eqnarray}
\nu p & \rightarrow & \mu^{-}p\pi^{+},\hspace{0.4cm}\nu n\rightarrow\mu^{-}p\pi^{0},\hspace{0.4cm}\nu n\rightarrow\mu^{-}n\pi^{+},\nonumber \\
\bar{\nu}n & \rightarrow & \mu^{+}n\pi^{-},\hspace{0.4cm}\bar{\nu}p\rightarrow\mu^{+}p\pi^{-},\hspace{0.4cm}\bar{\nu}p\rightarrow\mu^{+}n\pi^{0},\label{eq:process}
\end{eqnarray}
with $\nu$($\bar{\nu}$) energies exciting the second resonance region
and the corresponding cutoffs in $W_{\pi N'}$. We will obtain this
total cross section through the Eqs.(\ref{eq:Totalcross}-\ref{eq:EcmLab})with
the amplitude (\ref{eq:amplitude}), taking $(1-\gamma_{5})$ when
$\text{\ensuremath{\nu}}\rightarrow\bar{\nu}$, and the vertex production
contributions in Eqs.(\ref{eq:OBN}-\ref{eq:isospincoef}).

\subsection{Parameters and Form Factors}

What remains is to define the hadronic FF and the different coupling
constants. The coupling constant we use are the values from pion-nucleon
scattering, analysis of photo-production and electroproduction of
pions. For the strong couplings of nucleons we take $g_{_{\pi NN}}^{2}/4\pi=14$,(note
that $\frac{f_{\pi NN}}{m_{\pi}}=\frac{g_{\pi NN}}{2m_{N}}$) $g_{_{\rho NN}}^{2}/4\pi=2.9$,
$\kappa_{\rho}=3.7$, $g_{\omega NN}=3g_{_{\rho NN}}$ and $\kappa_{\omega}=-0.12$
\cite{Mariano01} with the usually adopted masses for involved hadrons
\cite{PTEP20,PDG06}. The coupling of nucleon $\rho$ and $\omega$
mesons were obtained by assuming the vector dominance model. In the
weak sector the vector coupling constant are fixed by assuming the
CVC hypothesis both, for B an R amplitudes. As usual, for the axial
currents we exploit the PCAC hypothesis and Golderberg-Treiman relations
with exception of the $\Delta$, the most important resonance in this
region, where the axial couplings are obtained by fitting to the differential
cross section(see below). For the nucleon Born and meson exchange
contributions in $\mathcal{O}_{BN}^{\lambda}$ we adopt $g_{_{V}}=1$,
$g_{_{\omega\pi V}}=0.324e$\cite{Mariano07}, while for the axial
couplings we assume $g_{_{A}}=1.26$ (PCAC values) and $f_{_{\rho\pi A}}=\frac{m_{_{\rho}}^{2}}{(93\text{MeV})g_{_{\rho NN}}}$
\cite{Sato03}.

The FF are expressed in terms of the usual Sachs dipole model for
the vector current and also a dipole FF for the axial part \cite{Barbero08}:
\begin{align}
F_{_{1}}^{V}(Q^{2}) & =\frac{g_{_{V}}}{1+t}\Bigg[G_{_{E}}^{p}(Q^{2})-G_{_{E}}^{n}(Q^{2})+t(G_{_{M}}^{p}(Q^{2})-G_{_{M}}^{n}(Q^{2}))\Bigg],\nonumber \\
F_{_{2}}^{V}(Q^{2}) & =\frac{g_{_{V}}}{1+t}\Bigg[G_{_{M}}^{p}(Q^{2})-G_{_{M}}^{n}(Q^{2})-(G_{_{E}}^{p}(Q^{2})-G_{_{E}}^{n}(Q^{2}))\Bigg],\nonumber \\
F^{A}(Q^{2}) & =\frac{g_{_{A}}}{(1+Q^{2}/M_{_{A}}^{2})^{2}},\hspace{0.5cm}M_{_{A}}=1.032\text{ GeV},\label{F12A}
\end{align}
where $t=Q^{2}/4m_{_{N}}^{2}$ and 
\[
G_{_{E}}^{p}(Q^{2})=\frac{1}{1+\kappa_{_{p}}}G_{_{M}}^{p}(Q^{2})=\frac{1}{\kappa_{n}}G_{_{M}}^{n}(Q^{2})=\frac{1}{1+Q^{2}/M_{_{V}}^{2}},\text{\hspace{0.5cm}}G_{_{E}}^{n}(Q^{2})=0,
\]
with $M_{_{V}}^{2}=0.71$ GeV$^{2}$, $\kappa_{_{p}}=1.79$, $\kappa_{_{n}}=-1.91$.
In the case of the contribution involving the $W\pi\pi$ vertex (third
term in Eq. \ref{eq:OBN}) we adopt the same $F_{_{V}}^{1}(Q^{2})$
as in the other Born terms (first, second and fourth terms in Eq.
(\ref{eq:OBN})) since these together should produce a gauge invariant
amplitude in the electromagnetic case.

Now, we define parameters and FF for the resonances. We begin with
the spin- $\frac{1}{2}$ ones.The coupling $f_{1440\pi N}$ can be
obtained from the partial decay width $N^{*}(1440)\rightarrow\pi N$
from Eq.(\ref{width1/2}) with $\pi=+1,m_{1440}=1462\text{ MeV},\Gamma_{1440}=391\text{ MeV}$\cite{PTEP20},
making the approach $\sqrt{s}\approx m_{1440}$ (CMW). We take $\Gamma_{_{1440\rightarrow\pi N}}\approx0.69\times391$
MeV $=269.79$ MeV \cite{Leitner09} were the factor $0.69$ comes
from the branching ratio of decay in $N\pi$ which is between 55\%
and 75\%. With this width we get the value of $f_{_{1440\pi N}}=0.412.$

The weak $W^{WN1440}$ vertex are obtained from the connection between
electromagnetic resonance production and the helicity amplitudes.The
helicity amplitudes describe the nucleon-resonances transition depending
on the polarization of the incoming photon and the spins of the baryons
\cite{Lakakulich06}. For non-zero $Q^{2}$, data on helicity amplitudes
for the $N^{*}(1440)$ are available only for the proton \cite{Lakakulich06},
where it is assumed that the isovector contribution on the neutrino
production is given as $g_{i}^{V}=-2g_{i}^{p}$. The PCAC hypothesis
allows us to relate the strong and weak FF and fix their values at
$Q^{2}=0$. We adopt the following parametrization and values taken
from Ref. \cite{Lakakulich06} 
\begin{eqnarray}
g_{1V}^{1440}(Q^{2})=\frac{4.6}{(1+Q^{2}/M_{V}^{2})^{2}(1+Q^{2}/4.3M_{V}^{2})} & , & g_{2V}^{1440}(Q^{2})=\frac{1.52}{(1+Q^{2}/M_{V}^{2})^{2}}(2.8\ln(1+Q^{2}/\text{GeV}^{2})-1),\nonumber \\
g_{1A}^{1440}(Q^{2}) & = & \frac{0.51}{(1+Q^{2}/M_{A}^{2})^{2}(1+Q^{2}/3M_{A}^{2})},\nonumber \\
g_{3A}^{1440}(Q^{2}) & = & 0.51\frac{(m_{1440}+m_{_{N}})}{Q^{2}+m_{\pi}^{2}}m_{N},\label{FF1440}
\end{eqnarray}
with $M_{V}=0.84$GeV and $M_{A}=1.05$ GeV. Note that the signs of
$g_{1V}(Q^{2}),g_{2V}(Q^{2}),g_{1A}(Q^{2})$ are the same that for
$F_{1V}(Q^{2}),F_{2V}(Q^{2}),F_{A}(Q^{2})$ in (\ref{F12A}) in spite
we have different form factors. For the $N^{*}(1535)$ resonance we
get from the same procedure followed before for the $N^{*}(1440)$
using Eq. (\ref{width1/2}) but for $\pi=-1,m_{1535}=1534\text{ MeV}$,
$\Gamma_{1535}=151\text{ MeV}$ and\textsf{\large{}{} }a branching-ratio
of 0.51 \cite{PDG06} a value{\large{} $f_{_{1535\pi N}}=0.17$, }while
for the weak FF we get\textsf{{} } 
\begin{align}
g_{1V}^{1535} & =\frac{4.0}{(1+Q^{2}/M_{V}^{2})^{2}(1+Q^{2}/1.2M_{V}^{2})}(7.2\ln(1+Q^{2}/\text{GeV}^{2})+1),\nonumber \\
g_{2V}^{1535} & =\frac{1.68}{(1+Q^{2}/M_{V}^{2})^{2}}(0.11\ln(1+Q^{2}/\text{GeV}^{2})),\nonumber \\
g_{1A}^{1535} & =\frac{0.21}{(1+Q^{2}/M_{A}^{2})^{2}(1+Q^{2}/3M_{A}^{2})},\nonumber \\
g_{3A}^{1535} & =\frac{0.21(m_{1535}-m_{_{N}})m_{_{N}}}{Q^{2}+m_{\pi}^{2}}.\label{eq:FF1535}
\end{align}
We analyze{\large{} }now the spin- $\frac{3}{2}$ resonances beginning
with the $\Delta$ one. For this resonance, for the mass, width and
$\pi N\Delta$ coupling constant we assume consistently the values
obtained previously from fitting the $\pi^{^{+}}p$ scattering data
\cite{Barbero08}, using the propagator (\ref{eq:G(-1/3)}) within
the CMS approach and the strong vertex (\ref{piNDeltavertex}). We
got $f_{_{N\pi\Delta}}^{2}/4\pi=0.317\pm0.003$, $m_{_{\Delta}}=1211.7\pm0.4$
MeV and $\Gamma_{_{\Delta}}=92.2\pm0.4$ MeV. For the vector $\Delta$
weak contribution to the B and R amplitudes we use the effective (empirical)
values $G_{_{M}}(0)=2.97$, $G_{_{E}}(0)=0.055$ and $G_{_{C}}(0)=\frac{2m_{_{\Delta}}}{m_{_{N}}-m_{_{\Delta}}}G_{_{E}}(0)$
fixed from photo and electroproduction reactions \cite{Mariano07,Sato1}.
We call these \textquotedbl{}effective\textquotedbl{} values, as discussed
in Ref. \cite{Mariano07}, because they correspond to the bare ones
$G_{_{i}}^{0}(0)$ (usually related with quark models(QM)) \textit{renormalized}
through the decay of a $\pi N$ state coming from the B amplitude
into a $\Delta$ through final state interactions (FSI). In Ref. \cite{Mariano07}
we also get the bare $G_{_{E,M}}^{0}(0)$ values by introducing dynamically
the FSI by an explicit evaluation of the rescattering amplitudes and
showed that the effective values, which are obtained through a fitting
procedure, can be in fact interpreted as the \textquotedbl{}dressed\textquotedbl{}
ones. For the FF we adopt 
\begin{eqnarray}
G_{_{i}}(Q^{2}) & = & G_{_{i}}(0)(1-Q^{2}/M_{_{V}}^{2})^{-2}(1+aQ^{2})e^{-bQ^{2}},\label{eq:GiQ2}
\end{eqnarray}
with $a=0.154$/GeV$^{2}$ and $b=0.166$/GeV$^{2}$, for $i=M$,
$E$, $C$, which corresponds also to Sachs dipole model times a corrections
factor already used in electroproduction calculations \cite{Sato1}.
The axial FF at $Q^{2}=0$, $D_{_{i}}(0),$ $i=1,4$, are obtained
by comparing the non-relativistic limit of the amplitude $\bar{u}_{_{\Delta}}^{\nu}A_{\nu\mu}u$
in the $\Delta$ rest frame ($p_{\Delta}=(m_{_{\Delta}},{\bf 0})$,
$p=(E_{_{N}}({\bf q}),$ $-{\bf q}$) with the non-relativistic QM
\cite{Sato03,Hol95}. $D_{4}(Q^{2})=0$ since we will not take into
account the contribution of the $\Delta$ deformation to the axial
current. The $Q^{2}$ dependence of $D_{i}$ is taken to be the same
as in vector case with a different parameter in the dipole factor,
i.e.

\begin{eqnarray}
D_{_{i}}(Q^{2}) & = & D_{i}(0)F(Q^{2}),\:\mbox{for }i=1,2,\hspace{0,5cm},D_{3}(Q^{2})=D_{3}(0)F(Q^{2})\frac{m_{_{N}}^{2}}{Q^{2}+m_{_{\pi}}^{2}},\label{FFADelta}
\end{eqnarray}
with $M_{_{A}}=1.02$ GeV and $F(Q^{2})=(1+Q^{2}/M_{A}^{2})^{-2}(1+aQ^{2})e^{-bQ^{2}}.$
Here

\[
D_{_{1}}(0)=\frac{3\sqrt{2}g_{_{A}}}{5}\frac{m_{_{N}}+m_{_{\Delta}}}{2m_{_{N}}F(-(m_{_{\Delta}}-m_{_{N}})^{2})},\,D_{2}(0)=-D_{1}(0)\frac{m_{N}^{2}}{(m_{N}+m_{\Delta})^{2}},\,D_{3}(0)=D_{1}(0)\frac{2m_{N}^{3}}{(m_{N}+m_{\Delta})m_{\pi}^{2}},
\]
where $F(-(m_{\Delta}-m_{N})^{2}$) in the denominator comes from
the fact that we scale $D_{i}(Q^{2}=-q^{2})$ from the time-like point
$q_{0}^{2}=(m_{\Delta}-m_{N})^{2}$ to $q^{2}=0$ through $F(Q^{2})$.
Then, as in the case of pion photo-production , we will consider $D_{1}(0)$
as a free (effective or empirical) parameter to be fitted from the
experimental data for $d\sigma/dQ^{2}$ and including the FSI effectively.
From this fit we get $D_{1}(0)=\frac{2.35}{\sqrt{2}}\mbox{ with }\chi^{2}/dof=0.71$,
and results are shown with full lines in the Fig.2 of Ref. \cite{Barbero08}
where a comparison with the data from the ANL and BNL experiments
\cite{Rad82,Kitagi86} of the neutrino flux $\phi(E_{\nu})$ averaged
cross section 
\begin{eqnarray*}
\left\langle \frac{d\sigma}{dQ^{2}}\right\rangle  & = & \frac{\int\limits _{E_{\nu}^{\text{min}}}^{E_{\nu}^{\text{max}}}\frac{d\sigma({E_{_{\nu}}})}{dQ^{2}}\phi(E_{\nu})dE_{\nu}}{\int\limits _{E_{\nu}^{\text{min}}}^{E_{\nu}^{\text{max}}}\phi(E_{\nu})dE_{\nu}},
\end{eqnarray*}
for the main channel $\nu p\longrightarrow\mu^{-}\pi^{+}p'$ with
the cut $W_{\pi N'}<1.4$ GeV in the final invariant mass is done.
With this cut it is expected, at less for this channel, that the contributions
of more energetic resonances than the $\Delta(1232)$ are small and
that are important for more energetic cuts. This will be analyzed
in the nect subsection. As the reanalyzed data of ANL achieved in
Ref.\cite{Rodriguez16} does not affect appreciably the channel used
to fit $D_{1}(0)$ for the mentioned cut we do not make a new fitting
with $\left\langle \frac{d\sigma}{dQ^{2}}\right\rangle $ nor show
again results for this, and we concentrate in the results for the
total cross section $\sigma(E_{\nu,\bar{\nu}})$ .

Now we fix parameters, coupling constant and FF for the $N^{*}(1520)$
resonance. From Eq.(\ref{Gamma1520}) making the CMW approach $\sqrt{s}\approx m_{1520}=1524$
MeV, $\Gamma_{1520}=115$ MeV and using the partial width $0.55$
\cite{PDG06} for decaying into $\pi N$ states we get $\frac{f_{\text{1520\ensuremath{\pi}N}}}{4\pi}=0.2$.
Choosing the values reported for the vector couplings for $Q^{2}=0$
in Ref.\cite{Leitner09}, we have for the vector part using the Eqs.(\ref{eq:CVGME})

\begin{align}
-2.98 & =\frac{{3}}{{2}}\frac{1.52}{0.94+1.52}(G_{M}^{1520}-G_{E}^{1520}),\nonumber \\
4.21 & =\frac{-{3}}{2}\Bigg(G_{M}^{1520}-\frac{4.56}{1.52-0.94}G_{E}^{1520}\Bigg)\frac{0.94}{0.94+1.52},
\end{align}
from where we get $G_{M}^{1520}=-2.62$,$G_{E}^{1520}=0.6$, while
for $\Delta$ was $G_{_{M}}^{\Delta}=2.97,G_{_{E}}^{\Delta}=0.055$,
being the change in $G_{M}$ consistent with the change of $C_{3}^{V}$
between both resonances (see Ref.(\cite{Lakakulich06})). If we use
$C^{V}$ values of Ref(\cite{Lakakulich06}) we get $G_{M}=-4.67$
and $G_{E}=-0.26$. For the axial couplings using the pole contribution
vertex we have using the $C^{A}$ values from Ref.\cite{Leitner09}

\begin{align}
D_{1}(0) & =-\frac{2.15}{\sqrt{2}},\nonumber \\
D_{2}(0) & =0,\nonumber \\
D_{3}(0) & =-\frac{2.15}{\sqrt{2}}\frac{m_{_{N}}^{2}}{m_{\pi}^{2}}.
\end{align}
while for $\Delta$ $D_{1}(0)=\frac{2.35}{\sqrt{2}}$, this is consistent
also with the change in $C_{A}^{5}$ (\cite{Lakakulich06}). The $Q^{2}$
dependence is assumed similar to that in Eqs. (\ref{eq:GiQ2}) and
(\ref{FFADelta}).

\subsection{Results}

We begin discussing a formal issue referred to spin- $\frac{3}{2}$
resonances. It would be useful to put attention in the Eq.(\ref{eq:propagator})
where the $\Delta$(also valid for the $N^{*}(1520)$ case) propagator
(\ref{propagA}) is written expanding the projectors in Eqs.(\ref{eq:G(-1/3)})
or Eqs.(\ref{eq:G(-1)}).

Note that if we take $A=-\frac{1}{3}$, our choice in previous works
\cite{Barbero08,Mariano07}, then $b(-\frac{1}{3})=2$ in (\ref{eq:propagator})
and $\frac{1}{2}(2Z+(1+4Z)A)=\frac{1}{3}Z-\frac{1}{6}$ in (\ref{Lint}).
At first, one could choose another value for $A$ while the same is
taken for the different vertexes coupled to the propagator. On the
other hand, if one takes $A=-1$ then $b(-1)=0$ and only the first
term of (\ref{propagA}), which sometimes is called (Wrongly) on shell
$\frac{3}{2}$ projector, contributes. Nevertheless, as can be seen
from (\ref{eq:G(-1/3)}), for a different value of $A$ $\frac{1}{2}$
off-shell propagation always is present. This is not property of the
$\frac{3}{2}$ field, also in the massive vector meson propagator
we have present an off-shell lower spin 0 component \cite{Berme89}.
As can be seen from the Eq.(\ref{eq:propagator}) for our election
the propagator has a contribution with a pole at $p^{2}=m_{\Delta}^{2}$
and another that is not singular. When the value $A=-1$ is adopted
this last term is not presented, nevertheless an observation regards
the vertexes should be done in order.

As was previously mentioned , quite generally in all interaction vertexes
we need a contact transformation invariant form proposed in (\ref{Lint}),
where $Z$ is an arbitrary parameter independent of $A$ that is property
of each interaction ( see Eq.(\ref{eq:Zparam})). We concentrate for
example in the strong $\pi N\Delta$ decay vertex for choosing $Z$,
while fix for simplicity the same value for the photo-production and
weak production ones as done in previous works \cite{Barbero08,Mariano07}.
Now we point to the question of the true degree of freedom of the
spin $\frac{3}{2}$ field, and remember that this is a constrained
quantum field theory. Observe that in the free RS Lagrangian in Eqs.(\ref{eq:Deltafree})
and (\ref{eq:Kinetic1}), there is no term containing $\dot{\Psi}_{0}$
. So, the equation of motion for it is a true constraint, and $\Psi_{0}$
has no dynamics. It is necessary then that interactions do not change
that fact and as it is shown in \cite{Badga17} this is fulfilled
for the value $Z=\frac{1}{2}$ . The same conclusion was obtained
in the original work of Nath \cite{Nath71}, where through other method
the same value was obtained. Then, we adopt this value for our interaction,
in resume we use $A=-1/3$ in propagators an vertexes involving the
$\Delta$ plus $Z=\frac{1}{2}$ being $\frac{1}{2}(1+4Z)A+Z=0$ and
$R_{\alpha}^{\mu}\left(0\right)=g_{\alpha}^{\mu}$ . This selection
will be the same for the $N^{*}(1520)$ that is an spin $\frac{3}{2}$
resonance. In spite of this analysis some authors \cite{Sato03,Nieves07,Rafi16,Lala10}try
to get both, the simpler versions for $\frac{3}{2}$ propagator using
$A=-1$ and a $\pi N\Delta$ vertex with $\frac{1}{2}(1+4Z)A+Z=0$.
This can be read in two different manners. First, if they are adopting
the same $Z=\frac{1}{2}$ value (generally this is not discussed at
all) as us, we could conclude that there is an inconsistence since
they are adopting a value $A=-1$ for the propagator while $A=-1/3$
to get $\frac{1}{2}(1+4Z)A+Z=0$, violating the independence of the
amplitude with $A$ . Or second, the different choice $Z=-\frac{1}{2}$
is adopted but not mentioned explicitly, and $A=-1$ it is used in
both propagator and vertexes. Nevertheless this $Z$ value does not
avoid the dynamics of $\Psi_{0}$ in the $\pi N\Delta$ vertex . En
each case, model dependencies are introduced. 

\textsf{\large{}{}}{\large{} }
\begin{figure}[h!]
\textsf{\large{}{}\includegraphics[width=20cm,height=16cm]{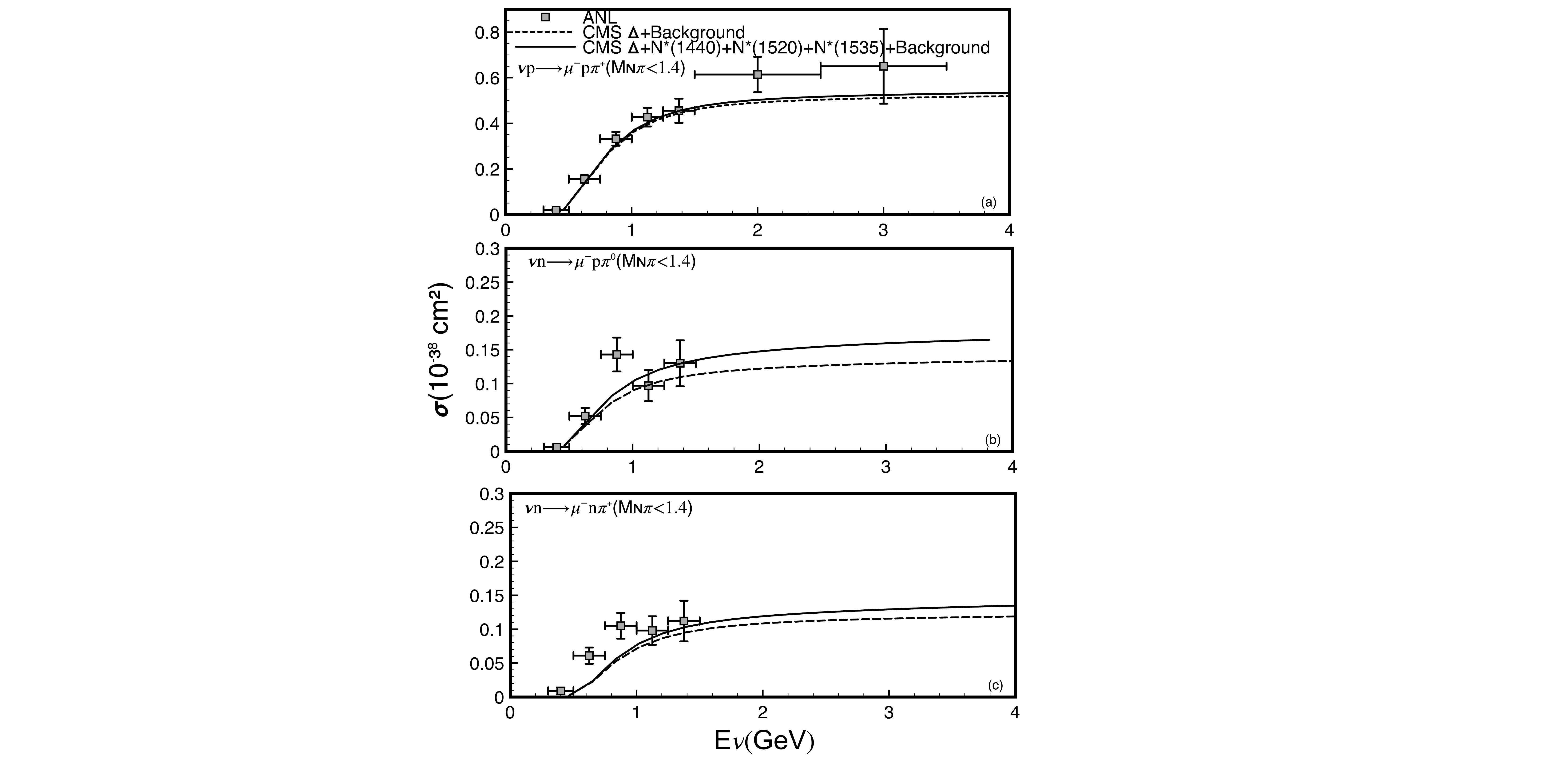}
\caption{Total $\nu N$ cross section as function of the neutrino energy for
different channel. Results with only the $\Delta$ and $\Delta+$
second region resonances plus the corresponding background, in each
case are shown for a cut $W_{\pi N}<1.4$ GeV. Data are taken form
Ref.\cite{Rad82}}
\label{cut1.4}}{\large{} }
\end{figure}
{\large \par}

\vspace{-0.5cm}
In Ref.\cite{Barbero08} we have showed the numerical consequences,
in the $\Delta$ region, of adopting the value $A=-1$ in the $\Delta$
propagator (called wrongly RS propagator in another works and referred
with this name there) keeping inconsistently $A=-1/3$ in the strong
and weak $\Delta$ vertexes for the value $Z=1/2$. Results are below
the consistent choice results and the data, showing that the inconsistence
leads to an observable effect.

Another formal problem is related with the fact that many works do
not consider the non-resonant background contributions \cite{Lakakulich06}
or do not introduce them through the corresponding effective Lagrangians,
or do not consider the interference between the background an resonant
contributions \cite{Lakakulich06,Leitner09} as really it is very
important to describe the data. On the other hand, also these models
detach the resonance production out of the weak production amplitude
\cite{Lakakulich06,Leitner09}. However, resonances are nonperturbative
phenomena associated to the pole of the S-matrix amplitude and one
cannot detach them from its production or decay mechanisms, being
necessary to built the amplitude through the Feynman rules using the
resonance propagators.

Now we show our results. Firstly, in Fig. (3) we compare our calculations
without and with the second resonance region included, for $W_{\pi N}<1.4$
GeV for the ANL data (BNL does not give results with this cut for
the total cross section). We implement the CMS approach for the $\Delta$
resonance, used previuosly in getting its strong and weak parameters
\cite{Barbero08,Mariano07,Mariano01}, and the CMW for the others.
As can be seen the effect of adding more resonances depends on the
considered channel. If we considered a fixed energy $E_{\nu}=3,1.5,1.5$
GeV for the mentioned $\nu$ channels in (\ref{eq:process}) respectively,
one can see from the Fig.(3) that their contribution is correspondingly
$4\%,17\%$ and $10\%$ , improving the data description regards the
model where one includes only the $\Delta$. Note that in spite of
the cut in $W_{\pi N}$, the tails of the resonances generated by
the finite width give an appreciable contribution and the interference
between them is also important.

Analyzing the individual contributions of the $N^{*}(1440),N^{*}(1520),$
and $N^{*}(1535)$ one can notice that the main contribution comes
from the $N^{*}(1520)$ being for the other less that $1\%$. \\
All isospin factors for the mentioned three isospin $\frac{1}{2}$
resonances read\medskip{}
\begin{eqnarray*}
\mathcal{T}_{h} & = & 0,\sqrt{2},-2,\mathcal{T}_{g}=-2,-\sqrt{2},0,
\end{eqnarray*}
and thus this explain why the contribution for the first channel is
small since comes from background terms of these resonances. For the
second channel we have the main effect since we have contributions
of both the direct and cross terms and interference between of them,
while for the last one we have only pole contribution\textsf{\large{}{}.
}In addition, one could to ask why the contribution of the second
resonance region are, apart from the cutting in invariant mass effect,
lower than the $\Delta$ + background contributions. This can be understood
from the Eqs.(\ref{eq:Totalcross}) and (\ref{eq:EcmLab}) . For certain
value of the neutrino energy $E_{\nu}$ in the Lab and $\nu N$ CM
systems $E_{\nu}^{CM}\sqrt{s}=E_{\nu}m_{N}$, being the limits in
the cross section integrals (\ref{eq:limits0}) fixed for a given
final $\mu\pi N'$ state, and if amplitudes are of the same order
of magnitude in the second resonance region regards the $\Delta$
one, the kinetical cross section factor $\frac{1}{E_{\text{\ensuremath{\nu}}}}$
favors smaller neutrinos energies and thus lowest excitation energy
contributions. For example if we take the final muon at rest $p_{R}^{2}=(E_{\nu}+m_{N}-m_{\mu})^{2}$
and thus for $p_{\Delta}^{2}=(1232)^{2}$ MeV$^{2}$ we have $\left(m_{N}E_{\nu}\right)^{-1}\approx2.7$GeV$^{-2}$
while for $p_{1520}^{2}=(1520)^{2}$ MeV$^{2}$ we have $\left(m_{N}E_{\nu}\right)^{-1}\approx1.5$GeV$^{-2}$.
Then, in spite strong and weak coupling constants would be of the
same order the $\Delta$ contribution is favored by the neutrino kinematical
factor. This explain the different size of the resonances contribution.

Information about the axial FF $D_{i}(0)$ for the $\Delta$ is carried
by the fitting to the differential cross section $d\sigma/dQ^{2}$
data for the cut $W_{\pi N}<1.4$GeV\cite{Barbero08} . Therefore,
contrasting the model predictions with the ANL and BNL differential
data cross sections, will help to complement the model's quality analysis.
Our results for the flux averaged cross sections are shown in Fig.
4, are shown for the $\Delta$ plus background, all resonances plus
background both coherent summed in the amplitude and with the incoherent
sum of resonant and background cross sections. As can be seen, the
efect of adding the second resonance region is noticiable and consistent
with the effect on the total cross section in the previous Fig. 3.
In addition, it is evident of the effect of adding resonant and backgroud
at the amplitud level (coherently) in place at the level of cross
section (incoherently).

\begin{figure}[t]
\includegraphics[width=9cm,height=8cm]{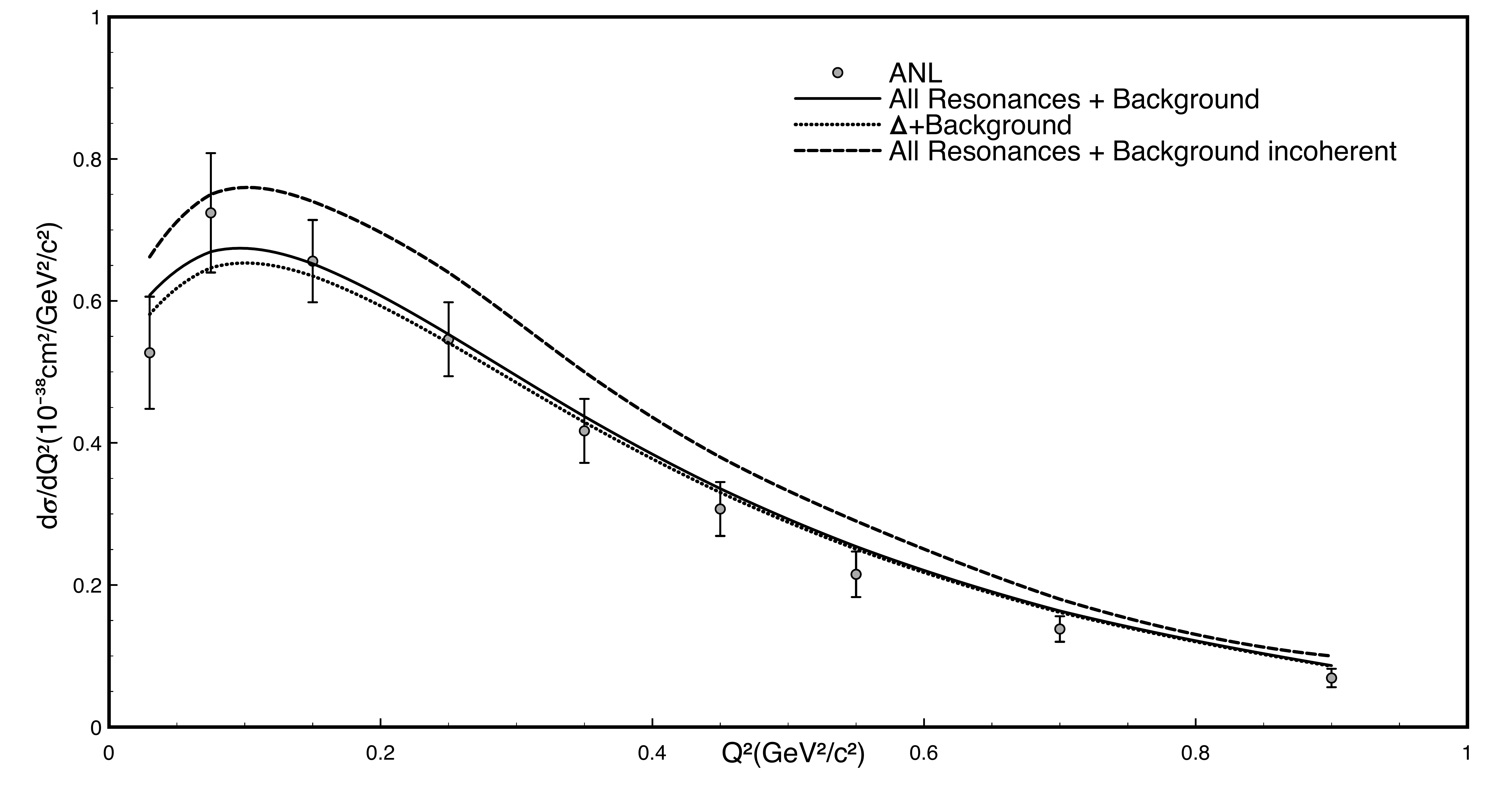}\includegraphics[width=9cm,height=8cm]{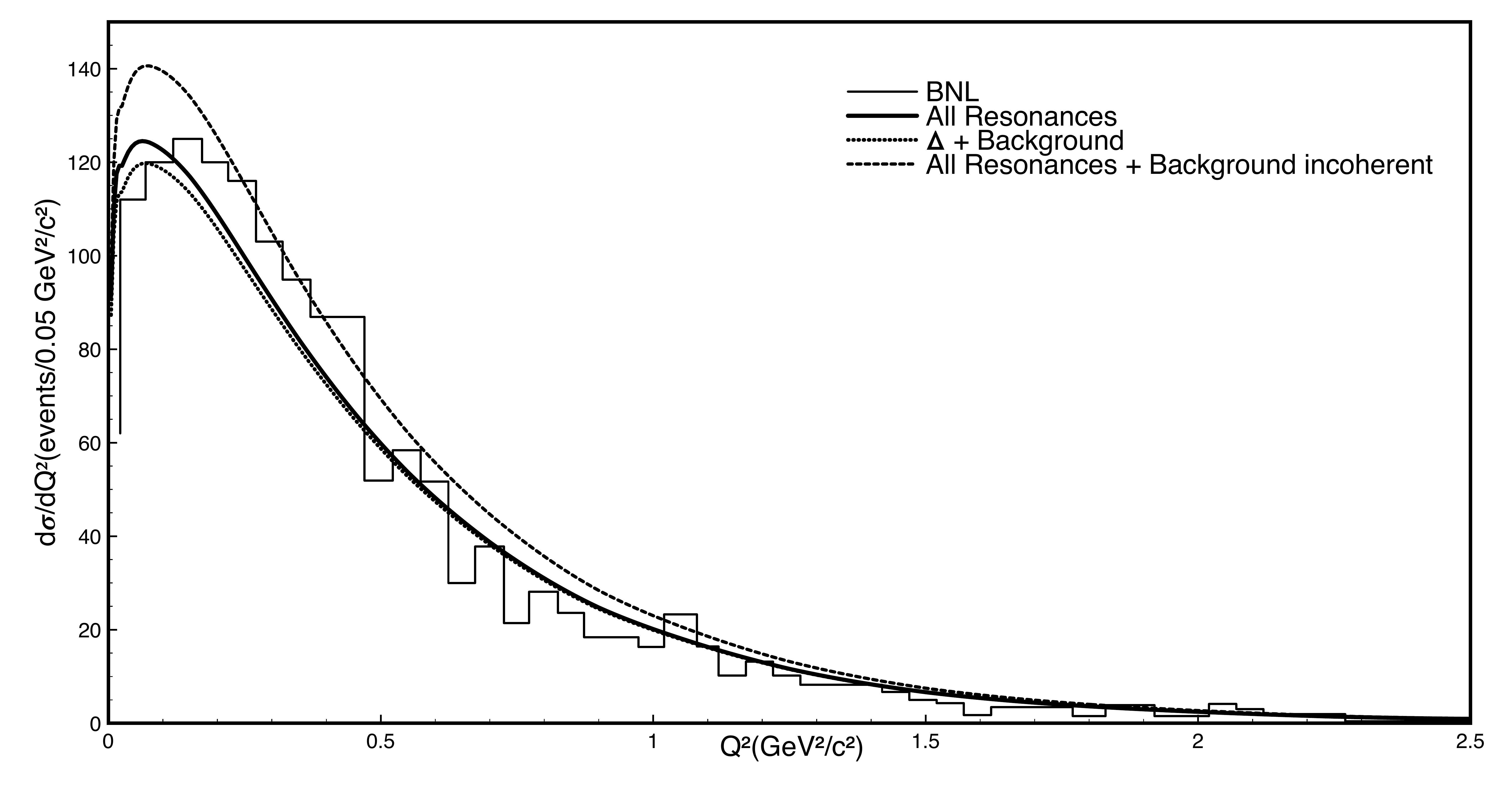}

\caption{Comparison of the calculated flux averaged differential cross section
$d\sigma/dQ^{2}$ for $W_{\pi N}<1.4$GeV with the data of referencies
\cite{Rad82,Kitagi86}. }
\end{figure}

\begin{figure}[H]
\textsf{\large{}{}\includegraphics[width=20cm,height=13cm]{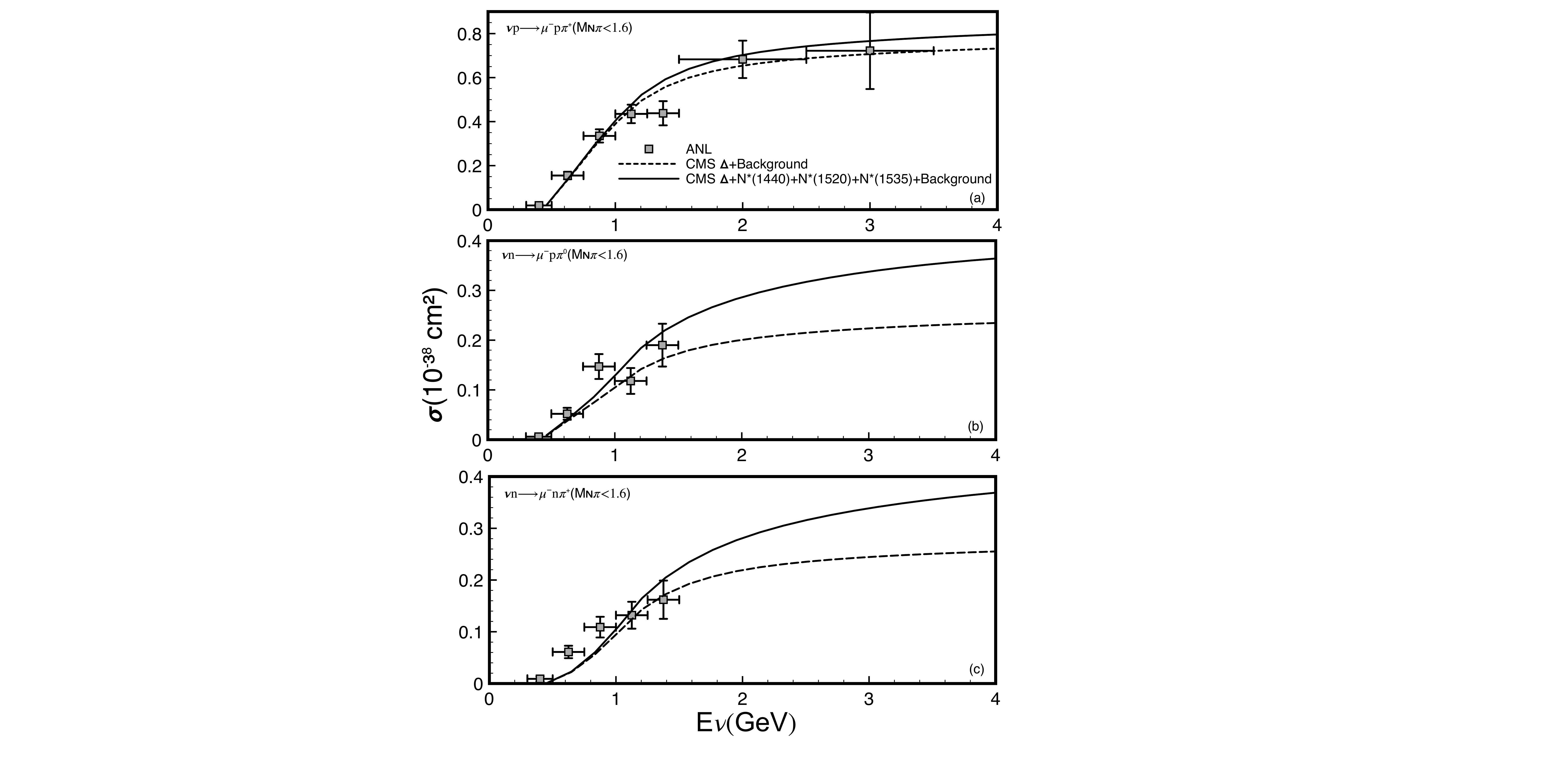}
\caption{Same as in Figure (\ref{cut1.4}) but for a cut $W_{\pi N}<1.6$ GeV.}
\label{cut1.6}}{\large{} }
\end{figure}

Now we go to the cut $W_{\pi N}<1.6$ GeV where the second resonance
region is fully included. As can be seen from the Fig.(\ref{cut1.6})
the contribution of these resonances is more important and necessary
to improve the consistence with data. Note that until this moment
we keep within the CMS and CMW approaches, the simplest to treat all
the resonances togheter the Born terms of the nonresonant background
that also include the resonance cross amplitudes.

\textsf{\large{}{}}{\large{} }
{\large \par}

Finally we compare the calculated flux averaged cross section $d\sigma/dW_{\pi N}$
for both cuts $1.4$ and $1.6$ GeV with the data for both ANL and
BNL experiments in Fig. 6, in order to see with more detail the contribution
of the resonant amplitud and background. As can be seen, we have a
large background contribution mainly in the $\nu n\rightarrow\mu^{-}\pi^{+}n$
channel, coming for the cross background contribution in Fig. 1(g)
for the $\Delta$. This contribution has isospin coefficient $-1$
while the resonant one in Fig. 1(h) has coefficient $-1/3$ giving
a small contribution to the cross section when squared. The responsable
of this behavior is the second term in the propagator (\ref{eq:propagator})
that is present, as consequence of our consistent selection of $A=-1/3,Z=1/2$
and grows for $p^{2}>m_{\Delta}^{2}$ . As this contribution cannot
be renormalyzed with a self energy as the pole Fig. 1(h) term, this
suggest the neccesity of FF for $W_{\pi N}$ to take into account
the finite size of the hadrons not considered in the puntual effective
vertexes\cite{Feuster98}. Of course, in another choices of $A,Z$
where $b(A)=0$ in (\ref{eq:propagator}) this growing is not present
but the treatment is not consistent. In addition, is not clear that
we can extend the another tree non-resonant background contributions
in Figs.1(a) to (f) to any final $W_{\pi N}$ keeping structureless
hadrons. Finally, it is visible the contribution of the $N^{*}(1520)$
resonance for the $\nu n\rightarrow\mu^{-}\pi^{+}n$ channel due to
the value of the isopin factor 2 in this case.

\begin{figure}
\includegraphics[width=10cm,height=11.5cm]{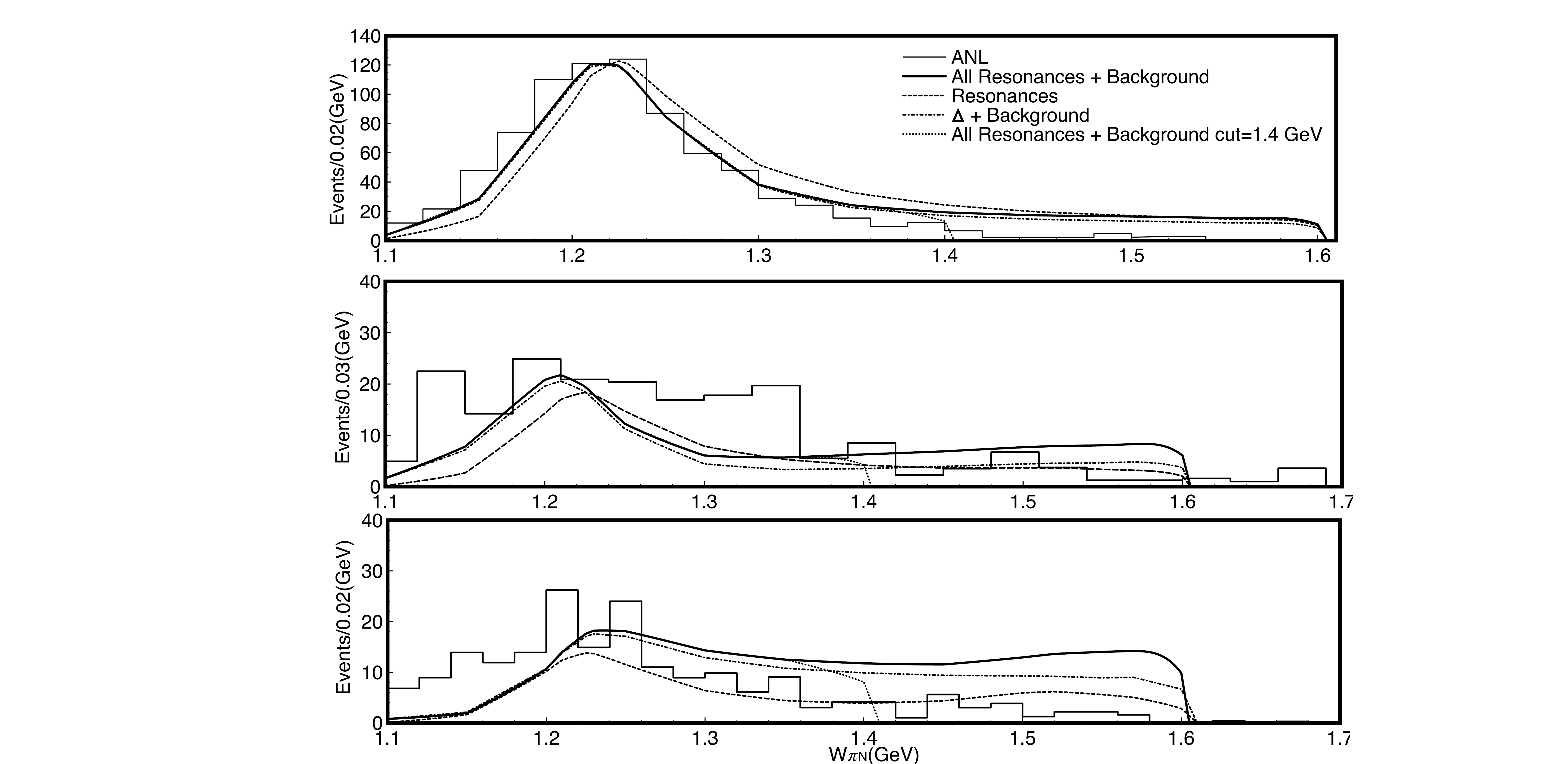}\negthickspace{}\negthickspace{}\negthickspace{}\negthickspace{}\negthickspace{}\negthickspace{}\negthickspace{}\negthickspace{}\negthickspace{}\negthickspace{}\negthickspace{}\negthickspace{}\negthickspace{}\negthickspace{}\negthickspace{}\includegraphics[width=9.5cm,height=11.1cm]{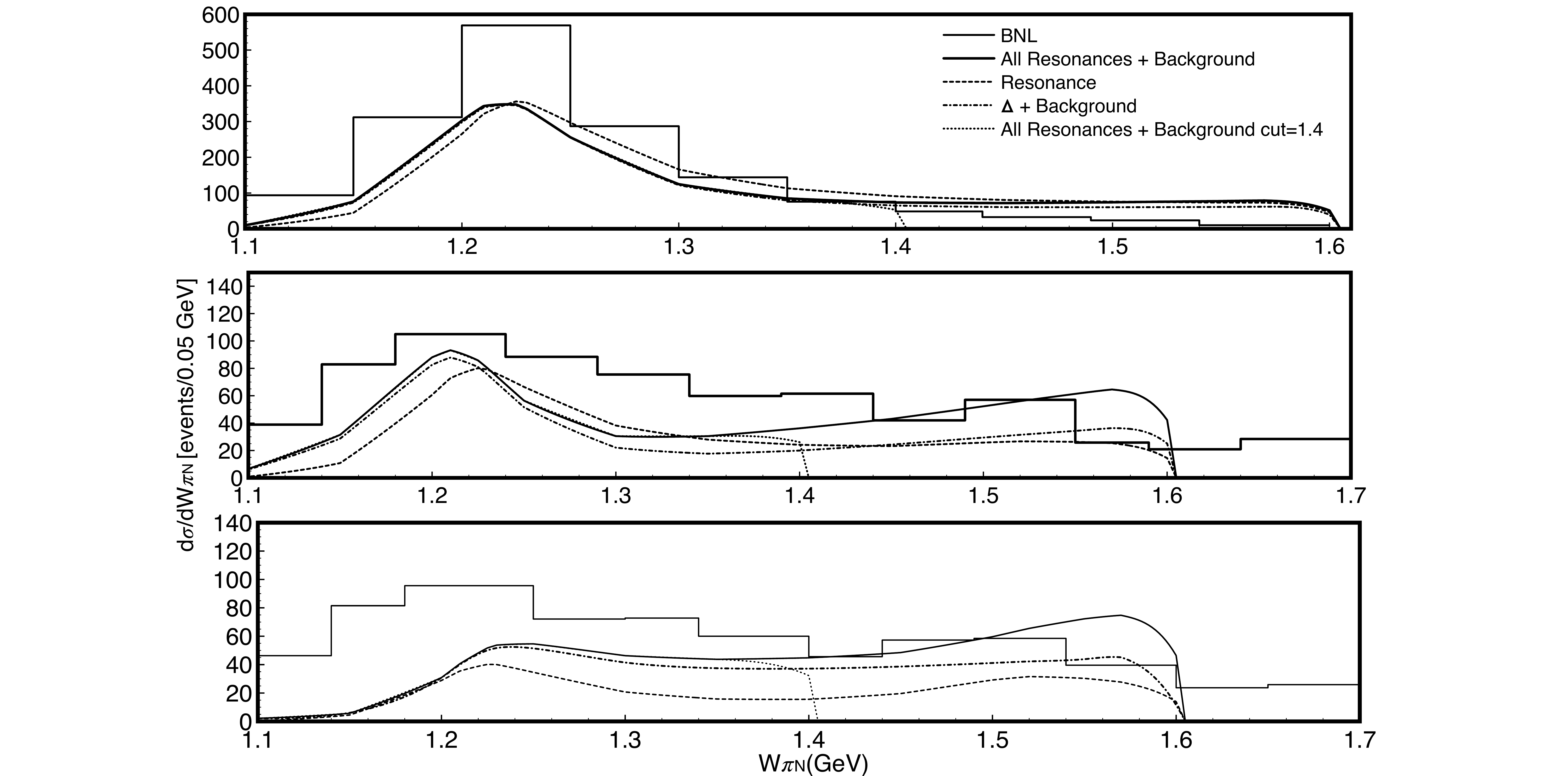}

\caption{Comparison of the flux averaged $d\sigma/dW_{\pi N}$ cross section
with the ANL and BNL data. The values of $W_{\pi N}$ for which theoretical
cross section is reported, correspond to the bin's central values
of the ANL or BNL data. }
\end{figure}

\begin{figure}[h!]
\textsf{\large{}{}\includegraphics[width=15cm,height=12cm]{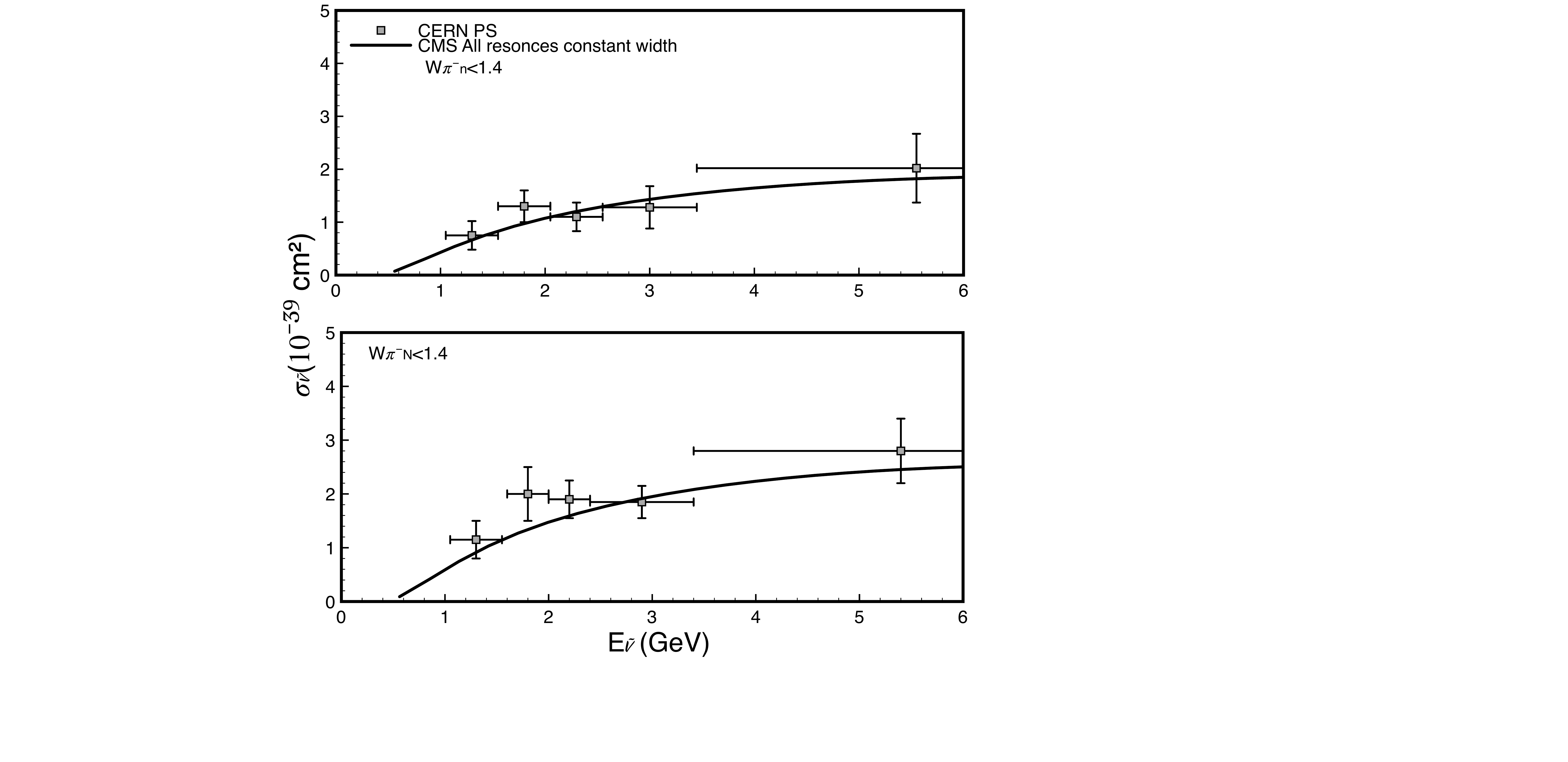}
\vspace{-2.5cm}
 }{\large \par}

\textsf{\large{}{}\caption{Antineutrino's total cross sections with a cut in $1.4$ GeV for the
$\bar{\nu}n\rightarrow\mu^{+}n\pi^{-}$ and that leading to a final
$N\pi^{-}$final state. }
\label{fig:Antinu1.4} }{\large \par}

\end{figure}

Now, in order to follow probing our model we wish to calculate the
antineutrinos total cross sections. We have two differences regards
the neutrinos case. Firstly, the interactions of neutrinos with hadrons
is not the same that for antineutrinos. We have a sign of difference
in the lepton current contraction that makes a different coupling
with the hadron one. Then, the interaction with neutrinos is different
from antineutrinos due the use of spinors for antiparticles in (\ref{eq:amplitude})
and has nothing to do with the very know CP violation effect. Secondly,
in the experiment we have an admixture of heavy freon CF3Br and was
exposed to the CERN PS antineutrino beam (peaked at $E_{\bar{\nu}}\sim$
1.5 GeV)\cite{Bolog79}. In this case the experiment informs that
we have 0.44\% on neutrons and 0.55\% of protons, and since our calculations
were for free nucleons we weight out results with these percentages
depending on the channel $\bar{\nu}n\rightarrow\mu^{+}n\pi^{-}$ or
$\bar{\nu}p\rightarrow\mu^{+}np\pi^{-}$. Our results including all
resonances and compared with the data are shown in Fig.(\ref{fig:Antinu1.4})
for the only cut $W_{\pi N}<1.4$ GeV reported in\cite{Bolog79} ,
and as can be seen we get a consistent description with the same cut
for the neutrino case.

\textsf{\large{}{}}{\large{} }
{\large \par}

Now we analyze the quality of our results and compare with other calculations
including the second resonance region, taking into account the formal
shortcomings mentioned above. As can be seen from a general point
of view, our model that fulfill consistence regards contact transformations
in the spin-$\frac{3}{2}$ field reproduce better the ANL data than
other inconsistent models \cite{Rafi16}. In addition, in that reference
it seems that the cross resonance contributions are omitted for the
$\nu p\rightarrow\mu^{-}\pi^{+}p$ channel. It is true that the direct
or pole contribution of isospin- $\frac{1}{2}$ resonances cannot
contribute to a isospin-$\frac{3}{2}$ amplitude, but the cross terms
do contribute noticing the isospin factors for this channel are not
zero in Eq.(\ref{eq:isospincoef}). This make the difference between
the full thin lines and the dashed ones in the upper panel of Figs.(\ref{cut1.4})
and (\ref{cut1.6}). A last shortcoming to mention is that for non
resonant backgrounds contributions Figs 1(a)-(f), an arbitrary cutoff
of $W_{\pi N}<1.2$ GeV is applied changing artificially the behavior
of these contributions independently from the rest of terms. This
is done for all the presented regimes $W_{\pi N}<1.4,1.6$ GeV. Note
that we can reproduce very well these data without the necessity of
any special cuttings, all contributions are calculated with the same
$W_{\pi N}$ maximum value.\\
 Finally, we note that in Ref.\cite{Rafi16} the antineutrino results
are not reproduced, while within our model the accordance with the
data is very well in all the energy region where the data is reported.

On the other hand, the model adopted in Ref.(\cite{Lakakulich06})
where the propagation of the resonance is described by a Breit-Wigner
distribution separating production and decay, does not include a background
amplitude and to get accordance in the data for the $\nu n\rightarrow\mu^{-}n\pi^{+},\hspace{0.4cm}\nu n\rightarrow\mu^{-}p\pi^{0}$
processes they needed to add incoherently a spin $\frac{1}{2}$ background.
The model adopted in Ref.(\cite{Leitner09}) is similar to that in
(\cite{Lakakulich06}) but they adjust the background cross section
contribution through a parameter $b^{\pi N}$ different for each channel.
These two last works were improved in Ref.\cite{Lala10}, where the
R and B contributions were added coherently and the $\Delta$ propagation
is treated with the choice $A=-1,Z=-1/2$ mentioned above, within
the parity conserving parametrization of the $WN\rightarrow\Delta$
vertex. These differences make difficult to compare with our results
when $R=\Delta$ since we choose a different $A,Z$ choice and the
Sachs parametrization. 

\section{Conclusions}

In this work we calculate the pion production cross section including
in the model spin- $\frac{1}{2}$ and $\frac{3}{2}$ resonances $\Delta(1232),N^{*}(1440),N^{*}(1520)$
and $N^{*}(1535)$ to cover the so called second resonance region.
From the formal point of view the spin- $\frac{3}{2}$ Lagrangians
( free and interaction) respect invariance under contact transformations
and the associated parameter $A$, is fixed to be the same in all
components of the Feynman rules to get $A$-independent amplitudes.
Also, the additional $Z$ parameter present in the $\mathcal{L}_{\pi NR}$
Lagrangian for these resonances is fixed to avoid time evolution of
the field component $\Psi_{0}$, since $\dot{\Psi}_{0}$ is not present
in the free Lagrangian. It is shown how another models do not analyze
these formal facts that can produced model dependence.

We treat the spin- $\frac{1}{2}$ resonances within the parity conserving
parametrization for the FF, since this is compatible to that used
in the similar topological nucleon contribution in Fig.1 (a) and (b).
For the spin- $\frac{3}{2}$ resonances we adopt the Sachs parametrization
to be consistent with our previous works including only the $\Delta$
resonance where we get better results than using the parity conserving
one \cite{Barbero14}. We followed the connection between both parameterizations
achieved in that reference to get the FF for the $N^{*}(1520)$ resonance
and have taken the $Q^{2}$ FF from Ref.(\cite{Lakakulich06}) for
all the second region resonances. We note that the main contribution
comes from the $\Delta$ resonance and the presence of the additional
resonances do not change appreciably the results of the cross section
for the $\nu p\rightarrow\mu^{-}p\pi^{+}$ channel for the cut $W_{\pi N}<1.4$
GeV as can be seen from Fig.(3). As this cut was used to fix $G_{M}(0),G_{E}(0)$
and $D_{1}(0)$ for the $\Delta$ we keep the same values obtained
previously in Refs.(\cite{Barbero08,Mariano07}). Firstly, we achieve
the comparison with the data of ANL experiment in the region of $W_{\pi N}<1.4,1.6$
GeV, where we have worked within the CMS+CMW approach. From the results
including and not including the second resonance region, we conclude
that to improve the data, this resonance region should be included.
More, in the case $W_{\pi N}<1.4$ GeV one can think why ? We conclude
this second energy region is necessary due to the tail of the resonances
that have their centroids out of this region, but influences through
the tails that interfere between resonance and background contributions.
This behavior is confirmed when we compare with the data with the
$W_{\pi N}<1.6$ GeV and the good agreement with the data for the
antineutrino case. In other approaches as in Ref.\cite{Hernandez17}
, the replacement ${\cal L}_{\pi N\Delta}\rightarrow{\cal L}_{\pi N\Delta}+c{\cal L}_{C}$
is proposed, with ${\cal L}_{C}$ describing contact terms without
the $\Delta$ field, with adjusting the low-energy constant c to get
a better fitting for the $\nu n\rightarrow\mu^{-}n\pi^{+}$channel.
The addition of contact terms is based on the argumentation that within
the ChPT framework, the equivalence between different Lagrangians
is at less of low energy constants, to be adjusted. We see that this
is not necessary if one includes consistently the second resonance
region.

Finally the data of Ref.(\cite{Rad82}) contains also results without
energy cuts and also all results in Ref.\cite{Kitagi86} are reported
without events exclusion. In addition a reanalysis of these two set
of data has been done recently in Ref.\cite{Rodriguez16} where the
main results are shown without cuts. For describing them we need to
extend the model to higher energies. This will be done in a next contribution.

\section{{\normalsize{}{}APPENDIX}}

\subsection{Spin projectors and isospin operators}

We have introduced $P_{ij}^{k}$ which {\small{}{}projects }on the
$k=\frac{3}{2}$, $\frac{1}{2}$ sector of the representation space,
with $i,j=1,2$ indicating the sub-sectors of the $\frac{1}{2}$ subspace,
and are defined as

\begin{eqnarray}
(P^{\frac{3}{2}})_{\mu\nu} & = & g_{\mu\nu}-\frac{1}{3}\gamma_{\mu}\gamma_{\nu}-\frac{1}{3p^{2}}\left[\ps\gamma_{\mu}p_{\nu}+p_{\mu}\gamma_{\nu}\ps\right],\nonumber \\
(P_{22}^{\frac{1}{2}})_{\mu\nu} & = & \frac{p_{\mu}p_{\nu}}{p^{2}},\nonumber \\
(P_{11}^{\frac{1}{2}})_{\mu\nu} & = & g_{\mu\nu}-P_{\mu\nu}^{\frac{3}{2}}-(P_{22}^{\frac{1}{2}})_{\mu\nu}\nonumber \\
 & = & (g_{\mu\alpha}-\frac{p_{\mu}p_{\alpha}}{p^{2}})(1/3\gamma^{\alpha}\gamma^{\beta})(g_{\beta\nu}-\frac{p_{\beta}p_{\nu}}{p^{2}}),\nonumber \\
(P_{12}^{\frac{1}{2}})_{\mu\nu} & = & \frac{1}{\sqrt{3}p^{2}}(p_{\mu}p_{\nu}-\ps\gamma_{\mu}p_{\nu}),\nonumber \\
(P_{21}^{1/2})_{\mu\nu} & = & \frac{1}{\sqrt{3}p^{2}}(-p_{\mu}p_{\nu}+\ps p_{\mu}\gamma_{\nu}).\label{eq:spinprojectors}
\end{eqnarray}
On the other hand we define the isospin $\Delta$ excitation operators

\begin{eqnarray*}
\boldsymbol{T}^{\dagger}\cdot\boldsymbol{\phi}_{+,-,0} & = & \left(\begin{array}{cc}
1 & 0\\
0 & \frac{1}{\sqrt{3}}\\
0 & 0\\
0 & 1
\end{array}\right),\left(\begin{array}{cc}
0 & 0\\
0 & 0\\
-\frac{1}{\sqrt{3}} & 0\\
0 & 1
\end{array}\right),\left(\begin{array}{cc}
0 & 0\\
\sqrt{\frac{2}{3}} & 0\\
0 & \sqrt{\frac{2}{3}}\\
0 & 0
\end{array}\right),
\end{eqnarray*}
that acts on $N=\left(\begin{array}{c}
1\\
0
\end{array}\right),\left(\begin{array}{c}
0\\
1
\end{array}\right)$ for proton and neutron respectively and 
\begin{eqnarray*}
\Delta_{++,+,0,-} & = & \left(\begin{array}{c}
1\\
0\\
0\\
0
\end{array}\right),\left(\begin{array}{c}
0\\
1\\
0\\
0
\end{array}\right),\left(\begin{array}{c}
0\\
0\\
1\\
0
\end{array}\right),\left(\begin{array}{c}
0\\
0\\
0\\
1
\end{array}\right)
\end{eqnarray*}
for the $\Delta$ states, being $\boldsymbol{\phi}{}_{+,-,0}=\frac{-1}{\sqrt{2}}(1,i,0),\frac{1}{\sqrt{2}}(1,-i,0),(0,0,1)$
and $\boldsymbol{W}_{\pm}=\boldsymbol{\phi}_{\pm}$.\\
 The isospin factors included in the resonances width are for isospin
$I=\frac{3}{2},\frac{1}{2}$ 
\begin{eqnarray*}
\Delta_{++}^{\dagger}\left(\boldsymbol{T}^{\dagger}\cdot\boldsymbol{\phi}_{+}\right)\left(\boldsymbol{T}\cdot\boldsymbol{\phi}_{+}^{\dagger}\right)\Delta_{++}=\Delta_{+}^{\dagger}\left[\left(\boldsymbol{T}^{\dagger}\cdot\boldsymbol{\phi}_{+}\right)\left(\boldsymbol{T}\cdot\boldsymbol{\phi}_{+}^{\dagger}\right)+\left(\boldsymbol{T}^{\dagger}\cdot\boldsymbol{\phi}_{0}\right)\left(\boldsymbol{T}\cdot\boldsymbol{\phi}_{0}^{\dagger}\right)\Delta_{+}\right] & =\cdots & =1\\
R^{\dagger}(\frac{1}{2})\left[\left(\tau\cdot\boldsymbol{\phi}_{0}\right)\left(\boldsymbol{\tau}\cdot\boldsymbol{\phi}_{0}^{\dagger}\right)+\left(\tau\cdot\boldsymbol{\phi}_{+}\right)\left(\boldsymbol{\tau}\cdot\boldsymbol{\phi}_{+}^{\dagger}\right)\right]R(\frac{1}{2}) & = & \cdots=3,
\end{eqnarray*}
since we can have $\pi^{0}p,\pi^{+}n,$ states when we have an isospin
$\frac{1}{2}$ projection ($q=e)$ and also $\pi^{0}n,\pi^{-}p$ when
isospin projection is $-\frac{1}{2}$ ($q=-e$) or $\pi^{-}n$ only
in the $I=3/2$ one with projection $-\frac{3}{2}$.

\subsection{Lagrangians and propagators involved in the non resonant background}

The propagators and interaction Lagrangians used to built amplitudes
$\mathcal{O}_{BN}$will be resumed here. First the propagators, which
come from the inversion of the kinetic operators present in the free
Lagrangians are 
\begin{eqnarray*}
S(p) & = & \frac{\ps+m_{N}}{p^{2}-m^{2}},\:\mbox{{nucleon}}\\
\Delta(p) & = & \frac{1}{p^{2}-m_{\pi}^{2}},\:\mbox{{pion}}\\
D_{\mu\nu}(p) & = & \frac{-g_{\mu\nu}+\frac{p_{\mu}p_{\nu}}{m_{V}^{2}}}{p^{2}-m_{V}^{2}},\:\mbox{{vector-meson}},
\end{eqnarray*}
while the effective strong interacting Lagrangians are

\begin{eqnarray*}
\mathscr{L}_{\pi NN}(x) & = & -\frac{g_{\pi NN}}{2m_{N}}\bar{\psi}(x)\gamma_{5}\gamma_{\mu}\boldsymbol{\tau}\cdot\left(\partial^{\mu}\boldsymbol{\boldsymbol{\phi}}(x)\right)\psi(x),\\
\mathscr{L}_{VNN}(x) & = & -\frac{g_{_{V}}}{2}\bar{\psi}(x)\left[\gamma_{\mu}\left\{ \begin{array}{c}
\boldsymbol{\rho}^{\mu}(x)\cdot\boldsymbol{\tau}\\
\omega^{\mu}(x)
\end{array}\right\} -\frac{\kappa_{V}}{2m_{N}}\sigma_{\mu\nu}\left(\partial^{\nu}\left\{ \begin{array}{c}
\boldsymbol{\rho}^{\mu}(x)\cdot\boldsymbol{\tau}\\
\omega^{\mu}(x)
\end{array}\right\} \right)\right]\psi(x),
\end{eqnarray*}
With $V=\omega,\rho$. Now, we define the effective hadron weak Lagrangians
built from Eqs.(\ref{eq:weakLagrangian},\ref{eq:hadroncurrent})

\begin{eqnarray*}
\mathscr{L}_{WNN}(x) & = & -\frac{g}{2\sqrt{2}}\overline{\psi}(x)\left[\gamma_{\mu}F_{1}^{V}(Q^{2})-\frac{F_{2}^{V}(Q^{2})}{2m_{N}}\sigma_{\mu\nu}\partial^{\nu}-F^{A}(Q^{2})\gamma_{\mu}\gamma_{5}\right]\sqrt{2}\boldsymbol{W}^{\mu}(x)\cdot\frac{\boldsymbol{\tau}}{2}\psi(x)+hc.,\\
\mathscr{L}_{W\pi\pi}(x) & = & -\frac{g}{2\sqrt{2}}F_{1}^{V}(Q^{2})\sqrt{2}\left[\boldsymbol{\boldsymbol{\phi}}(x)\times\partial_{\mu}\boldsymbol{\boldsymbol{\phi}}(x)\right]\cdot\boldsymbol{W}^{\mu}(x),\\
\mathscr{L}_{W\pi NN}(x) & = & -\frac{g}{2\sqrt{2}}\frac{f_{\pi NN}}{m_{\pi}}F_{1}^{V}(Q^{2})\bar{\psi}(x)\gamma_{5}\gamma_{\mu}\sqrt{2}(\boldsymbol{\tau}\times\boldsymbol{\boldsymbol{\phi}}(x))\cdot\boldsymbol{W}^{\mu}\psi(x),\\
\mathscr{L}_{W\pi\rho}(x) & = & \frac{g}{2\sqrt{2}}f_{\rho\pi A}F^{A}(Q^{2})\sqrt{2}\left(\boldsymbol{\boldsymbol{\phi}}(x)\times\boldsymbol{\rho}_{\mu}(x)\right)\cdot\boldsymbol{W}(x)^{\mu}\\
\mathscr{L}_{W\pi\omega}(x) & = & -\frac{g}{2\sqrt{2}}\frac{g_{\omega\pi V}}{m_{\omega}}F_{1}^{V}(Q^{2})\epsilon_{\mu\alpha\lambda\nu}\left(\partial^{\lambda}\boldsymbol{\boldsymbol{\phi}}(x)\right)\cdot\left(\partial^{\mu}\boldsymbol{W}{}^{\alpha}(x)\right)\omega^{\nu}(x).
\end{eqnarray*}

\section{Acknowledgments}

A. Mariano, belong to CONICET and UNLP, D.F. Tamayo Agudelo and D.E
Jaramillo Arango to UdeA.


\begin{thebibliography}{10}
\bibitem{Llewe72} C. H. Llewellyn Smith,Phys. Rep., \textbf{3}, 261,
(1972).

\bibitem[2]{Rad82} G. M. Radecky, et. al, Phys. Rev. D \textbf{25,}
1161 (1982).

\bibitem[3]{Kitagi86}T. Kitagaki,et al., Phys. Rev. D \textbf{34
}, 2554 (1986).

\bibitem[4]{Graczyk09}K.Graczyk,D.Kielczewska,P.Przewlocki,J.Sobczyk,Phys.Rev.
D 80, 093001 (2009). doi:10.1103/PhysRevD.80.093001

\bibitem[5]{Graczyk14}K.M. Graczyk, J. Zmuda, J.T. Sobczyk, Phys.
Rev. D 90, 093001 (2014). doi:10.1103/PhysRevD.90.093001

\bibitem[6]{wilki14}C. Wilkinson, P. Rodrigues, S. Cartwright, L.
Thompson, K. McFarland, Phys. Rev. D 90(11), 112017 (2014). doi:10.1103/
Phy

\bibitem[7]{Rodriguez16}Philip Rodrigues, Callum Wilkinson,and ,Kevin
McFarland, Eur. Phys. J. C (2016) 76:474.

\bibitem[8]{Kirbach2002}M. Kirchbach and D. Ahluwalia, Phys.Lett.\textbf{B},
(2002) 529124.

\bibitem[9]{Rarita}\textsf{\large{}{} }{\large{}W. Rarita and J.
Schwinger Phys. Rev.}\textbf{\large{} 60}{\large{}, 61 (1941).}{\large \par}

\bibitem[10]{Barbero08}\textsf{\large{}{} }{\large{}C. Barbero,
A. Mariano, G. Lopez Castro. Physics letters B 664 (2008) 70-77.}\textsf{\large{}{} }{\large \par}

\bibitem[11]{Mariano07}\textsf{\large{}{} }{\large{}A. Mariano.
Phys. lett. B (2007) 253; A. Mariano, J. Phys. G (2007) 1627.}{\large \par}

\bibitem[12]{Barbero12}C. Barbero, A. Mariano and G. L�pez Castro,
J. Phys. G: Nucl. Part. Phys. 39 (2012) 085011.

\bibitem[13]{Barbero15}C. Barbero, A. Mariano, J. Phys. G: Nucl.
Part. Phys. 42 (2015) 105104 .

\bibitem[14]{Amiri92}M. el Amiri, J. Pestieau and G. L\'{p}ez Castro,
Nucl. Phys. A 543( 1992) 673.

\bibitem[15]{PTEP20}P.A. Zyla et al. (Particle Data Group), Prog.
Theor. Exp. Phys. 2020, (2020)083C01.

\bibitem[16]{Lakakulich06}O. Lalakulich, E. A. Paschos, and G. Piranishvili,
Phys. Rev. D 74,(2006), 014009 .

\bibitem[17]{Leitner09} T. Leitner, O. Buss, L. Alvarez-Ruso, and
U. Mosel,\textsf{\large{}{} }\textit{\emph{\large{}Phys. Rev. C 79,(2009).}}{\large \par}

\bibitem[18]{Barbero14} C. Barbero, A. Mariano, G. Lopez Castro.
Physics Letters B 728 (2014) 282-287.

\bibitem[19]{Sato1}\textsf{\large{}{} }{\large{}T. Sato, T-S.H Lee,
Phys. Rev. C 63 (2001) 055201.}{\large \par}

\bibitem[20]{Mariano01}G. Lopez Castro, A. Mariano, Nucl. Phys. A
697 (2001)440.

\bibitem[21]{PDG06}Review of Particle Physics,W.-M. Yao, et al, J.
Phys. G 33 (2006) 1.

\bibitem[22]{Sato03}T. Sato, D. Uno, T.-S.H Lee, Phys. Rev. C 67
(2003) 065201.

\bibitem[23]{Hol95}T. R. Hemmert, B.R Holstein, Phys. Rev. D 51 (1995)
158.

\bibitem[24]{Berme89}M. Bermerrouche, R. M. Davidson, and N. C. Mukhopadhyay,
Phys. Rev. C 39, 2339 (1989).

\bibitem[25]{Bolog79}T. Bolognese, J.P. engel, J.L Guyonnet and J.L.
Riester, Phys. Lett. 81B (1979),393.

\bibitem[26]{Badga17}D. Badagnani, A. Mariano, and C. Barbero, J.
Phys. G: Nucl. Part. Phys. 44, 025001 (2017).

\bibitem[27]{Nath71}Nath L M, Etemadi B and Kimel J D 1971 Phys.
Rev. D 3 2153.\textsf{\large{}{} }{\large \par}

\bibitem[28]{Nieves07}E. Hernandez, J. Nieves, M. Valverde, Phys.
Rev. D 76, 033005 (2007). doi:10.1103/PhysRevD.76.033005

\bibitem[29]{Rafi16}M. Rafi Alam, M. Sajjad Athar, S. Chauhan and
S. K. SinghInternational Journal of Modern Physics E Vol. 25, No.
2 (2016) 1650010 .

\bibitem[30]{Lala10}O. Lalakulich, T. Leitner, O. Buss, and U. Mosel,
Phys. Rev. D 82, 093001 (2010)

\bibitem[31]{Feuster98}T. Feuster and U. Mosel, Phys. Rev. C 58,
457 (1998).

\bibitem[32]{Hernandez17}E. Hernandez and J. Nieves, Phys. Rev. D
95, 053007 (2017). 
\end{thebibliography}
\end{document}